\documentclass[%
 reprint,
unsortedaddress,
 amsmath,amssymb,
 aps,
]{revtex4-2}

\usepackage{hyperref}
\usepackage{physics}
\hypersetup{colorlinks=true, allcolors=blue}

\usepackage{graphicx}
\usepackage{dcolumn}
\usepackage{bm}
\usepackage{textcomp}
\usepackage{amsmath}
\def\mathclap#1{\text{\hbox to 0pt{\hss$\mathsurround=0pt#1$\hss}}}

\makeatletter
\newcommand{\vast}{\bBigg@{5}}
\newcommand{\Vast}{\bBigg@{5}}
\makeatother

\usepackage{xcolor}

\newcommand*{\colorboxed}{}
\def\colorboxed#1#{%
  \colorboxedAux{#1}%
}

\usepackage{tikz}
\usepackage{booktabs}
\usetikzlibrary{calc}

\newcommand*{\colorboxedAux}[3]{%
  \begingroup
    \colorlet{cb@saved}{.}%
    \color#1{#2}%
    \boxed{%
      \color{cb@saved}%
      #3%
    }%
  \endgroup
}

\usepackage{subfigure}

\usepackage{soul}

\usepackage{multirow}
\usepackage{xcolor}
\usepackage{soul}

\begin{document}

\preprint{APS/123-QED}

\title{Spin-Torque-driven Terahertz Auto Oscillations in Non-Collinear Coplanar Antiferromagnets}

\author{Ankit Shukla}
 \email{ankits4@illinois.edu}
\author{Shaloo Rakheja}
 \email{rakheja@illinois.edu}
\affiliation{Holonyak Micro and Nanotechnology Laboratory, University of Illinois at Urbana-Champaign, Urbana, IL 61801}

\date{\today}

\begin{abstract}
We theoretically and numerically study the terahertz auto oscillations, or self oscillations, 
in thin-film metallic non-collinear coplanar antiferromagnets (AFMs), such as $\mathrm{Mn_{3}Sn}$ and $\mathrm{Mn_{3}Ir}$, under the effect of antidamping spin torque with spin polarization perpendicular to the plane of the film. 
To obtain the order parameter dynamics in these AFMs, we solve three Landau-Lifshitz-Gilbert equations coupled by exchange interactions assuming both single- and multi-domain (micromagnetics) dynamical processes.
In the limit of a strong exchange interaction, the oscillatory
dynamics of the order parameter in these AFMs, which have opposite chiralities, could be mapped to that of two damped-driven pendulums \textcolor{black}{with significant differences in the magnitude of the threshold currents and the range of frequency of operation}. The theoretical framework allows us to identify the input current requirements as a function of the material and geometry parameters for exciting an oscillatory response. We also obtain a closed-form approximate solution of the oscillation frequency for large input currents in case of both $\mathrm{Mn_{3}Ir}$ and $\mathrm{Mn_{3}Sn}$. Our analytical predictions of threshold current and oscillation frequency agree well with the numerical results and thus can be used as compact models to design and optimize the auto oscillator.
Employing a circuit model, based on the principle of tunnel anisotropy magnetoresistance, we present detailed models of the output power and efficiency versus oscillation frequency of the auto oscillator.
Finally, we explore the spiking dynamics of two unidirectional as well as bidirectional coupled AFM oscillators using non-linear damped-driven pendulum equations.
Our results could be a starting point for building experimental setups to demonstrate auto oscillations in metallic AFMs, which have potential applications in terahertz sensing, imaging, and neuromorphic computing based on oscillatory or spiking neurons.
\end{abstract}

\maketitle

\section{Introduction}
Terahertz (THz) radiation, spanning from 100 Gigahertz (GHz) to 10 THz, are non-ionizing, have short wavelength, offer large bandwidth, scatter less, and are absorbed or reflected differently by different materials. 
As a result, THz electronics can be employed for safe biomedical applications, sensing, imaging, security, quality monitoring, spectroscopy, as well as for high-speed and energy-efficient non-von Neumann computing (e.g., neuromorphic computing).
THz electronics also has potential applications in beyond-5G communication systems and Internet of Things. Particularly, the size of the antennae for transmitting the electromagnetic signal could be significantly miniaturized in THz communication networks~\citep{tonouchi2007cutting, walowski2016perspective, mittleman2017perspective, son2019potential, ren2019state, rappaport2019wireless, elayan2019terahertz,  kurenkov2020neuromorphic}.
These aforementioned advantages and applications have led to an intense research and development in the field of THz technology with an aim to generate, manipulate, transmit, and detect THz signals~\citep{lewis2014review, mittleman2017perspective}. Therefore, the development of efficient and low power signal sources and sensitive detectors that operate in the THz regime is an important goal~\citep{mittleman2017perspective}. 

Most coherent THz signal sources can be categorized into three types --- particle accelerator based sources, solid state electronics based sources, and photonics based sources~\citep{mittleman2017perspective, lewis2014review}. Particle accelerator based signal generators include free electron lasers~\citep{tan2012terahertz}, synchrotrons~\citep{carr2002high}, and gyrotrons~\citep{idehara2008potential}. 
While particle accelerator sources have the highest power output, they require
a large and complex set-up~\citep{barh2015specialty}. 
Solid state generators include diodes~\citep{khalid2007planar, asada2016room, izumi20171}, 
transistors~\citep{otsuji2013emission, urteaga2017inp}, frequency multipliers~\citep{momeni2011broadband}, and Josephson junctions~\citep{kakeya2016terahertz}, whereas photonics based signal sources include quantum cascade lasers~\citep{williams2007terahertz}, gas lasers~\citep{chevalier2019widely}, and semiconductor lasers~\citep{kohler2002terahertz}. 
Solid-state generators are efficient at microwave frequencies whereas their output power and efficiency drop significantly above $100$ GHz~\citep{barh2015specialty}.
THz lasers, on the other hand, provide higher output power for frequencies above $30$ THz~\citep{pawar2013terahertz}, however, their performance for lower THz frequencies is plagued by noise and poor efficiency~\citep{barh2015specialty}. 
Here, we present the physics, operation, and performance benchmarks of a new type of nanoscale THz generator based on the ultra-fast dynamics of the order parameter of antiferromagnets (AFMs) when driven by spin torque.

Spin-transfer torque (STT)~\citep{slonczewski1996current, berger1996emission} and spin-orbit torque (SOT)~\citep{sinova2015spin} enable electrical manipulation of ferromagnetic order in emerging low-power spintronic radio-frequency nano-oscillators~\citep{chen2016spin}. 
When a spin current greater than a certain threshold (typically around $10^{8}-10^{9} \mathrm{A/cm^{2}}$~\citep{chen2016spin, khymyn2017antiferromagnetic}) is injected into a ferromagnet (FM) at equilibrium, the resulting torque due to this current pumps in energy which competes against the intrinsic Gilbert damping of the material. When the spin torque balances the Gilbert damping, the FM magnetization undergoes a constant-energy steady-state oscillation
around the spin polarization of the injected spin current. 
Such oscillators are nonlinear, current tunable with frequencies in the range of hundreds of MHz to a few GHz with output power in the range of nano-Watt (nW). They are also compatible with the CMOS technology~\citep{chen2016spin}; however, the
generation of the THz signal using FMs would require prohibitively large amount of current, which would lead to Joule heating and degrade the reliability of the electronics. It would also lead to electromigration and hence irreversible damage to the device set-up~\citep{kizuka2009dynamics}.

AFM materials, which are typically used to exchange bias~\citep{miltenyi2000diluted} an adjacent FM layer in spin-valves or magnetic tunnel junctions for FM memories and oscillators, have resonant frequencies in the THz regime~\citep{fiebig2008ultrafast, satoh2010spin, jungwirth2016antiferromagnetic, gomonay2018antiferromagnetic} due to their strong exchange interactions. 
It was suggested that STT could, in principle, be used to manipulate the magnetic order in conducting AFMs~\citep{gomonaui2008distinctive}, leading to either stable precessions for their use as high-frequency oscillators~\citep{gomonay2012symmetry, gomonay2014spintronics} or switching of the AFM order~\citep{gomonay2010spin} for their use as magnetic memories in spin-valve structures. 
The SOT-based spin Hall effect (SHE), on the other hand, could enable the use of both conducting~\citep{zarzuela2017antiferromagnetic, sulymenko2017terahertz, gomonay2018narrow, puliafito2019micromagnetic, lee2019antiferromagnetic} and insulating~\citep{cheng2016terahertz, khymyn2017antiferromagnetic, puliafito2019micromagnetic, parthasarathy2021precessional} AFMs in a bilayer comprising an AFM and a non-magnetic (NM) layer, for its use as a high frequency auto-oscillator~\citep{jenkins2013self}. 

\begin{figure}[ht!]
  \centering
  \includegraphics[width = 3.5in, clip = true, trim = 0mm 2mm 0mm 10mm]{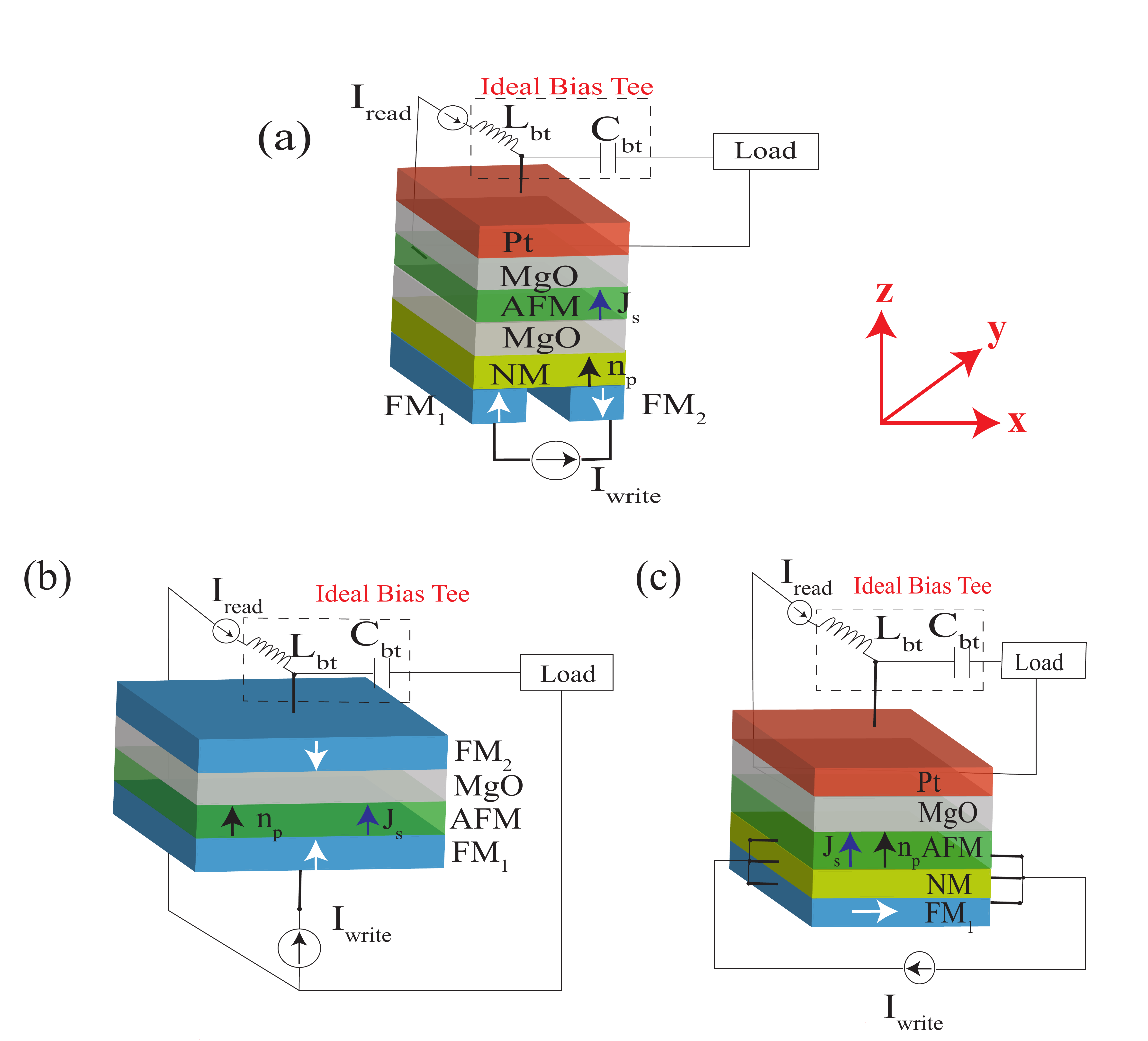}
  \caption{Device geometries to inject perpendicularly polarized spin current in thin-film metallic AFMs. In all the cases, $\mathrm{I_{write}}$ is the charge current injected to generate spin current, whereas, $\mathrm{I_{read}}$ is the charge current injected to extract the oscillations as a transduced voltage signal using the principles of tunnel anisotropy magnetoresistance (TAMR).
  (a) Lateral spin valve structure leads to spin accumulation in NM followed by injection into the AFM.
  (b) Perpendicular spin valve structure spin filters the injected charge current. 
  (c) FM/NM/AFM trilayer structure generates spin current due to interfacial spin-orbit torque.} 
  \label{fig:geometry}
\end{figure}

Table~\ref{tab:references} lists the salient results from some of the recently proposed AFM oscillators. 
These results, however, are reported \textcolor{black}{mainly} for collinear AFMs, while detailed analyses of the dynamics of the order parameter in the case of non-collinear AFMs is lacking. 
In this paper, firstly, we fill this existing knowledge gap in the modeling of auto oscillations in thin-film non-collinear coplanar AFMs like $\mathrm{Mn_{3}Ir}, \mathrm{Mn_{3}Sn}$, or $\mathrm{Mn_{3}GaN}$ under the action of a dc spin current. Secondly, we compare their performance (generation and detection) against that of collinear AFMs such as NiO for use as a THz signal source. In the case of NiO, inverse spin Hall effect (iSHE) is employed for signal detection, whereas in this work we utilize the large magnetoresistance of metallic AFMs. Finally, we investigate these auto oscillators as possible candidates for neuron emulators.
Considering that the spin polarization is perpendicular to the plane of the AFM thin-film, three possible device geometries are identified and presented in Fig.~\ref{fig:geometry} for the generation and detection of auto oscillations in metallic AFMs. 

\begin{table*}[t!]
\caption{Recent numerical studies on electrically controlled AFM THz oscillators. The investigated AFM materials, the direction of their uniaxial anisotropy axis $\vb{u}_{e}$ and that of the spin polarization of the injected spin current $\vb{n}_{p}$ are listed. Salient results along with the schemes to extract the oscillation as a voltage signal are also briefly stated. Ref.~\citep{lee2019antiferromagnetic} does not provide the name of a specific AFM, however, an AFM with uniaxial anisotropy is considered.}\label{tab:references}
\begin{ruledtabular}
\begin{tabular}{cllllr}
Ref.  &  AFM material & $\vb{u}_{e}$ & $\vb{n}_{p}$ & Salient Features & Detection Schemes \\
\hline
\citep{cheng2016terahertz} & $\mathrm{NiO}$ & $\vb{x}$ & $\vb{x}$ & a) THz oscillations for current above a threshold  & iSHE \\
& & & & b) Feedback in AFM/Pt bilayer sustains oscillation  &  \\
\citep{khymyn2017antiferromagnetic} &  {$\mathrm{NiO}$} & $\vb{x}$ & -$\vb{z}$ & a) Hysteretic THz oscillation in a biaxial AFM & iSHE \\
& & & & b) Threshold current dependence on uniaxial anisotropy &  \\
\citep{zarzuela2017antiferromagnetic} & $\alpha - \mathrm{Fe_{2}O_{3}}$ & a) $\vb{z}$ & $\vb{y}$ & a) Monodomain analysis of current driven oscillations in  & -\\
& & & & AFM insulators with DMI  &  \\
 &  & b) $\vb{y}$ & $\vb{y}$ & Similar to \citep{cheng2016terahertz} & -\\
\citep{sulymenko2017terahertz} &  $\alpha - \mathrm{Fe_{2}O_{3}}$ & $\vb{x}$ & $\vb{y}$ & a) Canted \textcolor{black}{net} magnetization due to DMI  & Dipolar radiation  \\
& & & & b) Small uniaxial anisotropy leads to low power THz frequency  &  \\
\citep{gomonay2018narrow} & $\mathrm{CuMnAs, Mn_{2}Au}$ & $\vb{y}$ &  $\vb{z}$ & a) Low dc current THz signal generation due to N\'eel SOT  & - \\
& & & & b) Phase locked detector for external THz signal  &  \\
\citep{puliafito2019micromagnetic}  & NiO & $\vb{x}$ & Varied & a) Comparison of analytical solutions to micromagnetic results & AMR/SMR  \\
& &  & & b) Effect of DMI on hysteretic nature of dynamics  &  \\
\citep{troncoso2019antiferromagnetic} & $\mathrm{Mn_{2}Ru_{x}Ga}$ & $\vb{z}$ & $\vb{y}$ & a) Generation of spin current in single AFM layer& AMR \\
& &  & & b) Oscillation dependence on reactive and dissipative torques  &  \\
\citep{lee2019antiferromagnetic} & Uniaxial Ani. &  $\vb{z}$ & Varied & a) Non-monotonic threshold current variation with $\vb{n}_{p}$ & - \\
& & & & b) Effects of anisotropy and exchange imperfections  &  \\
\citep{nomoto2020cluster} & {$\mathrm{Mn_{3}Sn}$} & x-y plane & $\vb{z}$ & a) Effective pendulum model based on multipole theory & AHE\\
\citep{parthasarathy2021precessional} & {$\mathrm{NiO, Cr_2O_3}$} & $\vb{x}$ & Varied & a) General effective equation of a damped-driven pendulum  & iSHE \\
& & & & b) Analytic expression of threshold current and frequency &  \\
Our & {$\mathrm{Mn_{3}Sn, Mn_{3}Ir}$} & x-y plane & $\vb{z}$ & a) Different numerical and analytic models   & TAMR \\
Work & & & & b) Inclusion of generation current for TAMR efficiency &  \\
& & & & c) Non-linear dynamics of bidirectional coupled oscillators &  \\
\end{tabular}
\end{ruledtabular}
\end{table*}

{Figure}~\ref{fig:geometry}(a) is based on the phenomena of spin injection and accumulation in a local lateral spin valve structure. Charge current, $\mathrm{I_{write}}$, injected into the structure is spin-polarized along the magnetization of $\mathrm{FM_{1}} (\mathrm{FM_{2}})$ and gets accumulated in the NM. It then tunnels into the AFM with the required perpendicular spin-polarization (adapted from Ref.~\citep{skarsvaag2015spin}). 
\textcolor{black}{The bottom MgO layer used here would reduce the leakage of charge current into the metallic AFMs considered in this work. This would reduce chances of Joule heating in the AFM thin-film layer.}
On the other hand, in Fig.~\ref{fig:geometry}(b), spin filtering~\citep{fujita2017field} technique is adopted wherein, a conducting AFM is sandwiched between two conducting FMs. Different scattering rates of the up-spins and down-spins of the injected electron ensemble at the two FM interfaces results in a perpendicularly polarized spin current as shown. 
The structure in Fig.~\ref{fig:geometry}(c) generates spin polarization perpendicular to the interface due to the interfacial SOTs generated at the FM/NM interface (adapted from Ref.~\citep{humphries2017observation, amin2020interfacial}). In this case, the spin current injected into the AFM has polarization along both $\vb{y}$ and $\vb{z}$ direction; however, the interface properties could be tailored to suppress the spin polarization along $\vb{y}$~\citep{humphries2017observation}.
In order to extract the THz oscillations of the order parameter as a measurable voltage signal, the tunnel anisotropy magnetoresistance (TAMR) measurements are utilized~\citep{park2011spin}.

In this work, we establish the micromagnetic model for non-collinear coplanar AFMs with three sublattices along with the boundary conditions in terms of both the sublattice magnetizations (Section~\ref{micro_theory}), as well as the N\'eel order parameter (Section~\ref{staggered}). 
We show that in the macrospin limit the oscillation dynamics correspond to that of a damped-driven pendulum (Section \ref{plus} and Section~\ref{minus}). 
The oscillation dynamics of AFM materials with two different chiralities in then compared in Section~\ref{comparison}.
We use the TAMR detection scheme to extract the oscillations as a voltage signal and present models of the output power and efficiency as a function of the oscillator's frequency (Section~\ref{extract}). This is followed by a brief investigation of the effect of inhomogeneity due to the exchange interaction on the dynamics of the AFM order (Section~\ref{micro}). 
Finally, we discuss the implication of our work towards building coherent THz sources in Section~\ref{disc}, and towards hardware neuron emulators for neuromorphic computing architecture in Section~\ref{appl}. Some of the salient results from this work are listed in Table~\ref{tab:references}.


\vspace{-5pt}
\section{Theory} \label{theory}
\vspace{-5pt}
\subsection{Magnetization Dynamics} \label{micro_theory}
We consider a micromagnetic formalism in the continuum domain~\citep{puliafito2019micromagnetic, ntallis2015micromagnetic} under which a planar non-collinear AFM is considered to be composed of three equivalent interpenetrating sublattices, each with a constant saturation magnetization $M_{s}$~\citep{yamane2019dynamics}. Each sublattice, $i \qty (= \text{1, 2 or 3})$, is represented as a vector field $\vb{m}_{i}(\vb{r}, t)$ such that for an arbitrary $\vb{r} = \vb{r}_{0}$, $\norm{\vb{m}_{i}(\vb{r}_{0}, t)} = 1$. The dynamics of the AFM under the influence of magnetic fields, damping, and spin torque is assumed to be governed by three Landau-Lifshitz-Gilbert (LLG) equations coupled by exchange interactions. For sublattice $i$, the LLG is given as~\citep{mayergoyz2009nonlinear}
\begin{equation}\label{eq:sLLGS}
    \begin{split}
       \pdv{\vb{m}_{i}}{t} &= - \gamma \mu_{0}\left(\vb{m}_{i} \cp \vb{H}_{i}^\mathrm{eff}\right) + \alpha_{i} \left(\vb{m}_{i} \cp \pdv{\vb{m}_{i}}{t}\right) \\  
        & \textcolor{black}{-} \omega_{s}\vb{m}_{i} \cp \left(\vb{m}_{i} \cp \vb{n}_{p}\right) \textcolor{black}{- \beta \omega_{s}\left(\vb{m}_{i} \cp \vb{n}_{p}\right)},
    \end{split}
\end{equation}
where $t$ is time in seconds, $\vb{H}_{i}^{\mathrm{eff}}$ is the position dependent effective magnetic field on $i$, $\alpha_{i}$ is the Gilbert damping parameter for $i$, and 
\begin{equation}
    \omega_{s} = \frac{\hbar}{2e}\frac{\gamma J_{s}}{M_{s} d_{a}}
\end{equation}
is the frequency associated with the input spin current density, $J_{s}$, with spin polarization along $\vb{n}_{p}$. Here, $d_{a}$ is the thickness of the AFM layer, $\hbar$ is the reduced Planck's constant, $\mu_{0}$ is the permeability of free space, $e$ is the elementary charge, and $\gamma = 17.6 \cp 10^{10} \mathrm{T^{-1} s^{-1}}$ is the gyromagnetic ratio. 
For all sublattices, the spin polarization, \textcolor{black}{$\vb{n}_{p}$}, is assumed to be along the ${z}$ axis. \textcolor{black}{Finally, $\beta$ is a measure of the strength of the field-like torque as compared to the antidamping-like torque. The effect of field-like torque on the sublattice vectors here is the same as that of an externally applied magnetic field---canting towards the spin polarization direction.
Results presented in the main part of this work do not include the effect of the field-like torque; however, a small discussion on the same is presented in the supplementary material~\citep{supple}}.

The effective magnetic field, $\mathbf{H}_{i}^{\mathrm{eff}}$ at each sublattice, includes contributions from internal fields as well as externally applied magnetic fields and is obtained as 
\begin{equation}\label{eq:h}
     \vb{H}_{i}^{\mathrm{eff}} \qty(\vb{r}, t) = -\frac{1}{\mu_{0} M_{s}}\fdv{\mathcal{F}}{\vb{m}_{i}\qty(\vb{r}, t)},
\end{equation}
where $\fdv{}{\vb{m}_{i}} = \pdv{}{\vb{m}_{i}} - \grad \vdot \pdv{}{\qty(\grad \vb{m}_{i})} $, and $\mathcal{F}$ is the energy density of the AFM, \textcolor{black}{considered in our work. It is} given as

\begin{flalign}\label{eq:energy_density}
    \begin{split}
        \mathcal{F} &= \sum_{\substack{\langle i, j \rangle}} \qty(\mathcal{J}\vb{m}_{i} \vdot \vb{m}_{j} + \mathcal{A}^{ij} \grad{\vb{m}_{i}} \vdot \grad{\vb{m}_{j}}) \\
                    &+ \mathcal{A}^{i i}\sum_{i =  1}^{3}\qty(\grad{\vb{m}_{i}})^{2} + \sum_{i = 1}^{3} \mathcal{K}_{h} m^{2}_{i, z} - \mathcal{K}_{e} \qty(\vb{m}_{i} \vdot \vb{u}_{e, i})^{2}\\
                    & + \mathcal{D} \sum_{\substack{\langle i, j \rangle}} \vb{z} \vdot \qty(\vb{m}_{i} \cp \vb{m}_{j}) + D^{ii}\sum \limits_{i = 1}^{3} \left(m_{i, z} \grad \vdot \vb{m}_{i} \right. \\
                    &\left.- \qty(\vb{m}_{i} \vdot \grad) m_{i, z}\right) \textcolor{black}{+ D^{ij}\sum_{\substack{\langle i, j \rangle}} \left( \left(m_{i, z} \grad \vdot \vb{m}_{j} \right. \right.} \\
                    &\textcolor{black}{\left. \left. - \qty(\vb{m}_{i} \vdot \grad) m_{j, z}\right) - \qty(m_{j, z} \grad \vdot \vb{m}_{i} - \qty(\vb{m}_{j} \vdot \grad) m_{i, z})\right)} \\
                    &- \sum_{i = 1}^{3}\mu_{0}M_{s}\vb{H}_{a} \vdot \vb{m}_{i},
    \end{split}
\end{flalign}
where $\langle i, j \rangle$ represents the sublattice ordered pairs $(1, 2)$, $(2, 3)$ and $(3, 1)$. 

The first three terms in Eq.~(\ref{eq:energy_density}) represent exchange energies. Here $\mathcal{J} (>0)$ is the homogeneous inter-sublattice exchange energy density whereas $\mathcal{A}^{i i} (>0)$ and $\mathcal{A}^{ij} (< 0)$ are the isotropic inhomogeneous intra- and inter-sublattice exchange spring constants, respectively. 
The next two terms in Eq.~(\ref{eq:energy_density}) represent magnetocrystalline anisotropy energy for biaxial symmetery upto the lowest order with $\mathcal{K}_{e} (>0)$ and $\mathcal{K}_{h} (>0)$ being the easy and hard axes anisotropy constants, respectively. We assume that the easy axes of sublattices 1, 2 and 3 are along $\vb{u}_{e, 1} = -(1/2) \vb{x} + (\sqrt{3}/2) \vb{y}$, $\vb{u}_{e, 2} = -(1/2) \vb{x} - (\sqrt{3}/2) \vb{y}$ and $\vb{u}_{e, 3} = \vb{x}$, respectively, and an equivalent out of plane hard axis exists along the $\vb{z}$ axis. 
The next three terms represent the structural symmetry breaking interfacial Dzyaloshinskii-Moriya Interaction (iDMI) energy density in the continuum domain. Its origin lies in the interaction of the antiferromagnetic spins with an adjacent heavy metal with a large spin-orbit coupling~\citep{thiaville2012dynamics, rohart2013skyrmion}. Here, we assume the AFM crystal to have $C_{nv}$ symmetry~\citep{bogdanov1989thermodynamically} such that 
the thin-film AFM is isotropic in its plane, and $\mathcal{D}$, $D^{ii}$, and $D^{ij}$ represent the effective strength of homogeneous and inhomogeneous iDMI, respectively, along the $\vb{z}$ direction. Finally, the last term in Eq.~(\ref{eq:energy_density}) represents the Zeeman energy due to an \textcolor{black}{externally applied} magnetic field $\vb{H}_{a}$. 
\textcolor{black}{Now,} using Eq.~(\ref{eq:energy_density}) in Eq.~(\ref{eq:h}) we get the effective field for sublattice $i$ as
\begin{flalign}\label{eq:h2}
    \begin{split}
        \vb{H}_{i}^{\mathrm{eff}}  &= \sum_{\substack{j \\ j \neq i}}\qty(-\frac{\mathcal{J}}{\mu_{0} M_{s}} \vb{m}_{j} + \frac{\mathcal{A}^{ij}}{\mu_{0} M_{s}} \grad^{2}{\vb{m}_{j}}) + \frac{2\mathcal{A}^{i i}}{\mu_{0} M_{s}}\grad^{2}{\vb{m}_{i}} \\
                                   &- \frac{2\mathcal{K}_{h}}{\mu_{0} M_{s}} m_{i, z} \vb{z} + \frac{2\mathcal{K}_{e}}{\mu_{0} M_{s}} \qty(\vb{m}_{i} \vdot \vb{u}_{e, i}) \vb{u}_{e, i}\\
                                   &+\frac{\mathcal{D} \vb{z} \cp \qty(\vb{m}_{j} - \vb{m}_{k})}{\mu_{0} M_{s}} -\frac{2D^{ii}}{\mu_{0} M_{s}} \left(\qty(\grad \vdot \vb{m}_{i}) \vb{z} - \grad m_{i, z}\right) \\
                                   &\textcolor{black}{-\frac{D^{ij}}{\mu_{0} M_{s}} \left(\qty(\grad \vdot \qty(\vb{m}_{j} - \vb{m}_{k})) \vb{z} - \grad (m_{j, z} - m_{k, z})\right)} \\
                                   &+ \vb{H}_{a},
    \end{split}
\end{flalign}
where $(i, j, k) = (1, 2, 3), (2, 3, 1),$ or $(3, 1, 2)$, respectively. 

In order to explore the dynamics of the AFM, we adopt a finite difference discretization scheme and discretize the thin-film of dimension $L \times W \times d_{a}$ into smaller cells of size $s_{L} \times s_{W} \times s_{d}$. Each of these cells is centered around position $\vb{r}$ such that $\vb{m}_{i} (\vb{r}, t)$ denotes the average magnetization of the spins within that particular cell~\citep{puliafito2019micromagnetic}. Finally, we substitute Eq.~(\ref{eq:h2}) in Eq.~(\ref{eq:sLLGS}) and use fourth-order Runge-Kutta rule along with the following boundary conditions for sublattice $i$ of the thin-film considered (see supplementary material~\citep{supple}):
\begin{flalign}\label{eq:bc1}
    \begin{split}
        &2\mathcal{A}^{i i}\pdv{\vb{m}_{i}}{\pmb{\eta}} + \mathcal{A}^{ij} \sum_{\substack{j \\ j \neq i}}\vb{m}_{i} \cp \qty(\pdv{\vb{m}_{j}}{\pmb{\eta}} \cp \vb{m}_{i}) \\
        &+ D^{ii} \vb{m}_{i} \cp \qty(\pmb{\eta} \cp \vb{z}) \\
        &\textcolor{black}{+ {D^{ij}} \vb{m}_{i} \cp \qty(\vb{m}_{i} \cp \qty(\qty(\pmb{\eta} \cp \vb{z}) \cp \qty(\vb{m}_{k} - \vb{m}_{j})))} = 0,
    \end{split}
\end{flalign}
where $\pmb{\eta}$ is the normal vector perpendicular to a surface parallel to $\vb{x}$ or $\vb{y}$. \textcolor{black}{The above equation ensures that the net torque due to the internal fields on the boundary magnetizations of each sublattice is zero in equilibrium as well as under current injection~\citep{abert2019micromagnetics}. For the energy density presented in Eq.~(\ref{eq:energy_density}), the fields at the boundary are non-zero only for inhomogeneous inter- and intra-sublattice exchange, and Dzyaloshinskii-Moriya interactions. 
Finally,} in the absence of DMI, we have the Neumann boundary condition $\pdv{\vb{m}_{i}}{\pmb{\eta}} = 0$, which implies that the boundary magnetization does not change along the surface normal $\pmb{\eta}$. 

For all the numerical results presented in this work we solve the system of Eqs.~(\ref{eq:sLLGS}),~(\ref{eq:h2}), and~(\ref{eq:bc1}) with the equilibrium state as the starting point. The equilibrium solution in each case was arrived at by solving these three equations for zero external field and zero current with a large Gilbert damping of $0.5$. 


\subsection{N\'eel Order Dynamics}\label{staggered}
The aforementioned micromagnetic modeling approach assuming three sublattices is extremely useful in exploring the physics of the considered AFM systems. It is, however, highly desirable to study an effective dynamics of the AFMs under the effect of internal and external stimuli in order to gain fundamental insight. Therefore, we consider an average magnetization vector $\vb{m}$ and two staggered order parameters $\vb{n}_{1}$ and $\vb{n}_{2}$ to represent an equivalent picture of the considered AFMs. These vectors are defined as~\citep{gomonay2012symmetry, gomonay2015using, yamane2019dynamics}
\begin{subequations}\label{eq:staggered}
    \begin{flalign}
        &\vb{m} \equiv \frac{1}{3} \left(\vb{m}_{1} + \vb{m}_{2} + \vb{m}_{3}\right),\label{eq:m1} \\
        &\vb{n}_{1} \equiv \frac{1}{3\sqrt{2}} \left(\vb{m}_{1} + \vb{m}_{2} - 2\vb{m}_{3}\right), \label{eq:n1}  \\
        &\vb{n}_{2} \equiv \frac{1}{\sqrt{6}} \left(-\vb{m}_{1} + \vb{m}_{2}\right), \label{eq:n2} 
    \end{flalign}
\end{subequations}
such that $\norm{\vb{m}}^{2} + \norm{\vb{n}_{1}}^{2} + \norm{\vb{n}_{2}}^{2} = 1$. 
The energy landscape (Eq.~(\ref{eq:energy_density})) can then be represented as
\begin{flalign}\label{eq:energy_density2}
    \begin{split}
        \frac{\mathcal{F}}{3} &= \frac{3\mathcal{J}}{2}\vb{m}^{2} + \mathcal{A}_{m}\qty(\grad{\vb{m}})^{2} + \frac{\mathcal{A}_{n}}{2}\qty(\qty(\grad{\vb{n}_{1}})^{2} + \qty(\grad{\vb{n}_{2}})^{2}) \\
                    &+ \mathcal{K}_{h}\left(m^{2}_{z} +  n^{2}_{1, z} + n^{2}_{2, z}\right) - \frac{\mathcal{K}_{e}}{2} \left(\frac{3}{2} \left(n_{1, x} - n_{2, y}\right)^2 \right. \\
                    &\left. + \frac{1}{2}\qty(n_{1, y} + n_{2, x})^2 + m_{x}\qty(m_x - \sqrt{2}\qty(n_{1, x} - n_{2, y}))\right. \\
                    &\left. + m_y\qty(m_{y} +\sqrt{2} \qty(n_{1, y} + n_{2, x})) + 4n_{1, x} n_{2, y} \right) \\
                    &+ \sqrt{3}\mathcal{D} \vb{z} \vdot \qty(\vb{n}_{1} \cp \vb{n}_{2}) + D^{ii} \left(m_{z} \grad \vdot \vb{m} - \qty(\vb{m} \vdot \grad) m_{z} \right.\\
                    &+ \left.n_{1, z} \grad \vdot \vb{n}_{1} - \qty(\vb{n}_{1} \vdot \grad) n_{1, z} + n_{2, z} \grad \vdot \vb{n}_{2} \right.\\
                    &- \left. \qty(\vb{n}_{2} \vdot \grad) n_{2, z} \right) \textcolor{black}{+ \sqrt{3}D^{ij} \left(n_{1, z} \grad \vdot \vb{n}_{2} - \qty(\vb{n}_{1} \vdot \grad) n_{2, z} \right.} \\
                    &\textcolor{black}{-\left. n_{2, z} \grad \vdot \vb{n}_{1} + \qty(\vb{n}_{2} \vdot \grad) n_{1, z} \right)} - \mu_{0}M_{s}\vb{H}_{a} \vdot \vb{m}, 
    \end{split}
\end{flalign}
where $\mathcal{A}_{m} = \left(\mathcal{A}^{i i} + \mathcal{A}^{i j}\right)$ and $\mathcal{A}_{n} = \left(2\mathcal{A}^{i i} - \mathcal{A}^{i j}\right)$. 
\begin{figure*}[ht!]
  \centering
  \includegraphics[width = 2\columnwidth, clip = true, trim = 0mm 2mm 0mm 0mm]{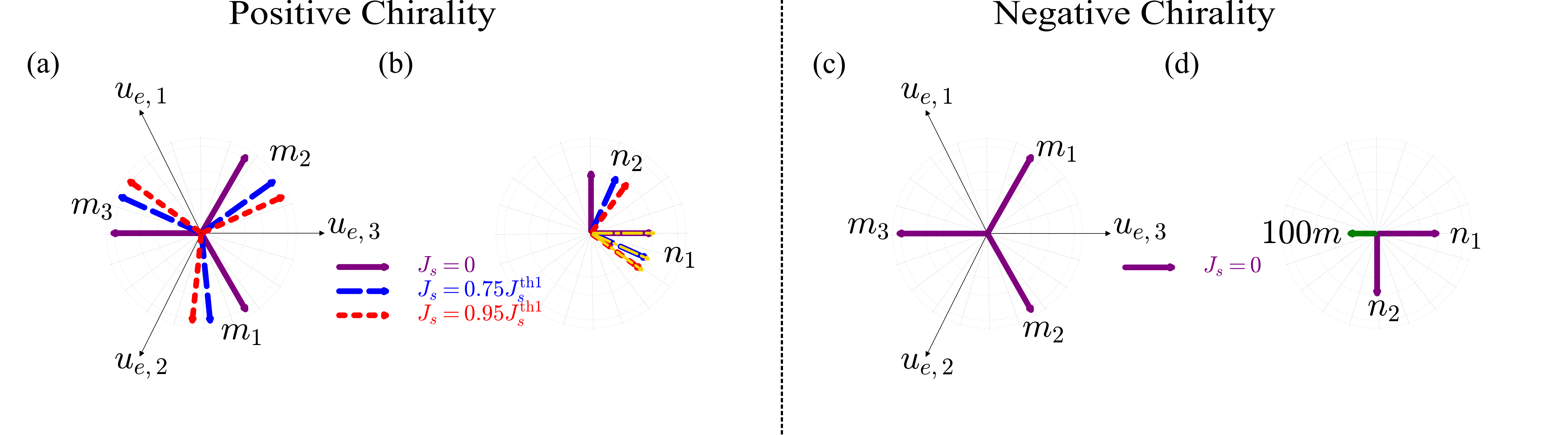}
  \caption{Stationary solution for AFMs with positive (a, b) and negative chirality (c, d). 
  (a) Sublattice magnetization for currents below the threshold current $J_{s}^{\mathrm{th1}}$. When $J_{s} = 0$ (equilibrium state), the sublattice vectors $\vb{m}_{i}$ coincide with the easy axes $\vb{u}_{e, i}$, whereas for a non-zero current smaller than $J_{s}^{\mathrm{th1}}$, the macrospins have stationary solutions other than the equilibrium solution, as depicted by dashed and dotted line. The $z$ component of these vectors is zero. 
  (b) An equivalent representation of (a) through the staggered order parameters $\vb{n}_{1}$ and $\vb{n}_{2}$. They are perpendicular to each other \textcolor{black}{and have zero out-of-plane component}. The thinner dash-dotted gold arrows through the thicker arrows of $\vb{n}_{1}$ represent the analytic expression of the stationary solution $\varphi = -\frac{1}{2}\sin^{-1}\left({\frac{2\omega_{s}}{\omega_{K}}}\right)$. 
  \textcolor{black}{The average magnetization $\vb{m}$ is vanishingly small, as can also be noticed from Eq.~(\ref{eq:m2}).}
  (c) Sublattice magnetization at equilibrium ($J_{s} = 0$). The $z$ component of these vectors is zero. Here, only the sublattice vector $\vb{m}_{3}$ coincides with its corresponding easy axis. On the other hand, $\vb{m}_{1}$ and $\vb{m}_{2}$ are oriented such that the energy due to DMI is dominant over anisotropy. 
  (d) An equivalent representation of (c) through the staggered order parameters $\vb{n}_{1}$ and $\vb{n}_{2}$, and average magnetization $\vb{m}$. $\vb{n}_{1}$ and $\vb{n}_{2}$ are almost perpendicular to each other with a negative chirality, as assumed in Eq.(\ref{eq:n1_n2}). A small in-plane net magnetization (shown by the magnified green arrow) also exists in this case~\citep{yamane2019dynamics}.} 
  \label{fig:stationary}
\end{figure*}

An equation of motion involving the staggered order parameters can be obtained by substituting Eq.~(\ref{eq:sLLGS}) in the first-order time derivatives of Eq.~(\ref{eq:staggered}) and evaluating each term carefully (see supplementary material~\citep{supple}). However, an analytical study of such an equation of motion that consists of contributions from all the energy terms of Eqs.~(\ref{eq:energy_density}) or~(\ref{eq:energy_density2}) would be as intractable as the dynamics of individual sublattices itself. Therefore, we consider the case of AFMs with strong inter-sublattice exchange interaction \textcolor{black}{such that $\mathcal{J} \gg |\mathcal{D}| \gg \mathcal{K}_e $}. \textcolor{black}{This corresponds to systems with ground state confined to the easy-plane ($\vb{x-y}$ plane) and those that host $\norm{\vb{m}}\ll 1$ (weak ferromagnetism), $\vb{n}_{1}$ $\perp$ $\vb{n}_{2}$, and $\norm{\vb{n}_{1}} \approx \norm{\vb{n}_{2}} \approx 1/\sqrt{2}$~\citep{gomonay2015using, yamane2019dynamics}.} 
\textcolor{black}{However, when an input current is injected in the system, the sublattice vectors cant towards the spin polarization direction leading to an increase in the magnitude of $\vb{m}$ while decreasing that of $\vb{n}_1$ and $\vb{n}_2$. Spin polarization along the $\vb{z}$ direction and an equal spin torque on each sublattice vector ensures that $\vb{n}_1$ and $\vb{n}_2$ have negligible $\vb{z}$ components at all times (Eqs.~(\ref{eq:n1}),~(\ref{eq:n2})).} 
Therefore, we consider 
\begin{subequations}\label{eq:n1_n2}
    \begin{flalign}
    \vb{n}_{1}(\vb{r}, t) &= \lambda \begin{pmatrix}
                                \sqrt{1-n_{1z}^{2}}\cos \varphi(\vb{r}, t) \\
                                \sqrt{1-n_{1z}^{2}}\sin \varphi(\vb{r}, t) \\
                                n_{1z}(\vb{r}, t)
                                \end{pmatrix},\\
    \vb{n}_{2}(\vb{r}, t) &= \lambda \begin{pmatrix}
                                \sqrt{1-n_{2z}^{2}}\cos (\varphi(\vb{r}, t) \pm \pi/2) \\
                                \sqrt{1-n_{2z}^{2}}\sin (\varphi(\vb{r}, t) \pm \pi/2) \\
                                n_{2z}(\vb{r}, t)
                                \end{pmatrix},
\end{flalign}
\end{subequations}
where $\varphi$ is the azimuthal angle from the $x$ axis and $|n_{1z}|, |n_{2z}| \ll 1$. 

The two choices for $\vb{n}_{2}$ correspond to two different classes of materials---one with a positive ($+\pi/2$) chirality and the other with a negative ($-\pi/2$) chirality~\citep{yamane2019dynamics}. Materials that have a negative (positive) value of $\mathcal{D}$ correspond to $+\pi/2 (-\pi/2)$ chirality because the respective configuration reduces the overall energy of the system. $\mathrm{L1_{2}}$ phase of AFMs like $\mathrm{Mn_{3}Ir}, \mathrm{Mn_{3}Rh}$, or $\mathrm{Mn_{3}Pt}$ is expected to host $+\pi/2$ chirality whereas the hexagonal phase of AFMs like $\mathrm{Mn_{3}Sn}, \mathrm{Mn_{3}Ge}$, or $\mathrm{Mn_{3}Ga}$ is expected to host $-\pi/2$ chirality~\citep{vzelezny2017spin, yamane2019dynamics}.

\begin{figure*}[ht!]
  \centering
  \includegraphics[width = 2\columnwidth, clip = true, trim = 0mm 2mm 0mm 0mm]{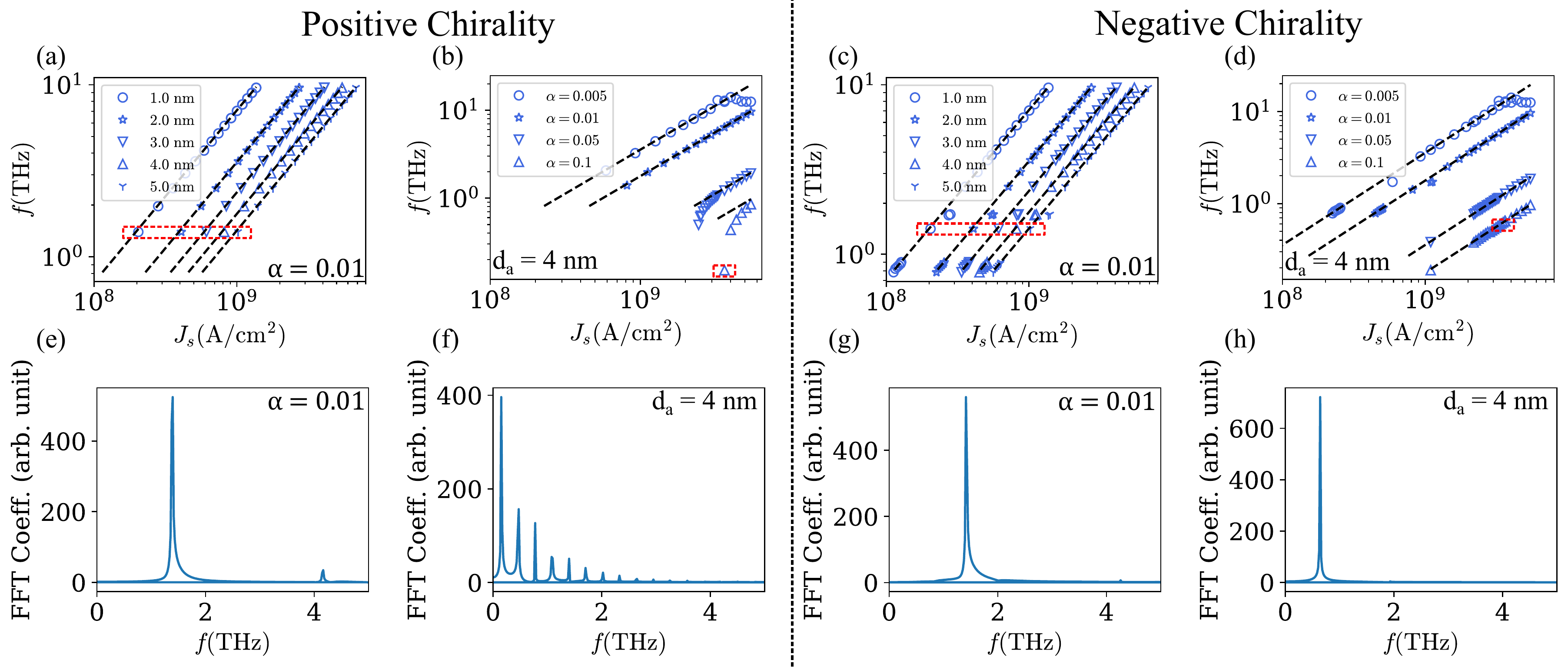}
  \caption{Upper panel (a-d) shows the time averaged frequency as a function of input spin current, whereas the lower panel (e-h) shows the FFT of the oscillations corresponding to the cases marked by the dashed red boxes above. In the time averaged frequency response in the upper panel, the dashed black lines denote the analytic expression of frequency (Eq.~(\ref{eq:frq})). 
  (a), (c) Frequency response for different film thicknesses for $\alpha = 0.01$. 
  (e), (g) FFT of the signal corresponding to $J_{s} = J_{s}^{\mathrm{th2}}$. 
  (b), (d) Frequency response for different damping constants for $d_{a} = 4 \ \mathrm{nm}$. 
  (f), (h) FFT of the signal corresponding to $\alpha = 0.1$ and $J_{s}^{\mathrm{th1}}$, respectively.
  Positive chirality: The numerical values of the average frequency match very well against the analytic expression for lower damping and large current. On the other hand, non-linearity and, hence, higher harmonics are observed for small current and large damping.
  Negative chirality: The numerical values of the average frequency exactly match against the analytic expression for all values of damping and input current considered here.} 
  \label{fig:freq}
\end{figure*}  

\section{Single Domain Analysis}
\subsection{Positive Chirality}\label{plus}
Defining $\vb{n}_{3} = \qty(\vb{n}_{1} \cp \vb{n}_{2})/{\lambda}$, and considering the case of $+\pi/2$ chirality, it can be shown that $\vb{m}$ is just a dependent variable of the N\'eel order dynamics. To a 
first order, $\vb{m}$ could be expressed as~\citep{gomonay2015using, yamane2019dynamics} 
\begin{equation}\label{eq:m2}
    \begin{split}
        \vb{m} =& - \frac{1}{\omega_{E}}\left(\vb{n}_{1} \cp \vb{\dot{n}}_{1} + \vb{n}_{2} \cp \vb{\dot{n}}_{2} + \vb{n}_{3} \cp \vb{\dot{n}}_{3} - \gamma \mu_{0} \vb{H}_{a} \right.\\
        &\left. \textcolor{black}{- \beta \omega_{s} \vb{n}_{p}}\right),
    \end{split}
\end{equation}
where $\omega_{E} = 3\gamma \mathcal{J}/M_{s}$. One can then arrive at the equation of motion for the N\'eel vectors as
\begin{equation}\label{eq:motion1}
    \begin{split}
        &\vb{n}_{1} \cp \left[\vb{\ddot{n}}_{1} - c^{2} \grad^{2}{\vb{n}_{1}} - \omega_{E} \boldsymbol{\omega}_{Kn1} + \omega_{E} {\omega}_{K_{h}} \qty(\vb{n}_{1} \vdot \vb{z}) \vb{z} \right. \\
        &+ \left. \omega_{E} \omega_{\mathcal{D}}(\vb{n}_{2} \cp \vb{z}) + \omega_{E}\boldsymbol{\omega}_{Dn1}^{ii} + \omega_{E}\boldsymbol{\omega}_{Dn2}^{ij} + \alpha \omega_{E} \dot{\vb{n}}_{1} \right.\\
        &-\left. \omega_{E}\omega_{s}\qty(\vb{n}_{1} \cp \vb{n}_{p})\right] + \vb{n}_{2} \cp \left[\vb{\ddot{n}}_{2} - c^{2} \grad^{2}{\vb{n}_{2}} - \omega_{E} \boldsymbol{\omega}_{Kn2}\right. \\
        &+ \left. \omega_{E} {\omega}_{K_{h}}\qty(\vb{n}_{2} \vdot \vb{z}) \vb{z} - \omega_{E} \omega_{\mathcal{D}}(\vb{n}_{1} \cp \vb{z}) + \omega_{E}\boldsymbol{\omega}_{Dn2}^{ii}\right. \\
        &- \left. \omega_{E}\boldsymbol{\omega}_{Dn1}^{ij} + \alpha \omega_{E} \dot{\vb{n}}_{2} - \omega_{E}\omega_{s}\qty(\vb{n}_{2} \cp \vb{n}_{p})\right] + \vb{n}_{3} \cp \ddot{\vb{n}}_{3} \\
        &+ \gamma \mu_{0} \left(\vb{n}_{1} \cp \dot{\vb{n}}_{1} + \vb{n}_{2} \cp \dot{\vb{n}}_{2} + \vb{n}_{3} \cp \dot{\vb{n}}_{3}\right) \cp \vb{H}_{a} \\
        &- \gamma \mu_{0} \dot{\vb{H}}_{a} = 0,
    \end{split}
\end{equation}
where $c = \sqrt{\omega_{E} \gamma \mathcal{A}_{n}/M_{s}}$, 
$\omega_{K_h} = 2\gamma \mathcal{K}_h/M_s$, 
$\boldsymbol{\omega}_{K, n1} = \frac{\omega_K}{4} \left(\qty(n_{1, y} + n_{2, x} + \sqrt{2}m_y) \vu{x} +  \left(n_{1, x} + 3n_{2, y} \right. \right.\\ 
\left. \left. +\sqrt{2} m_x \right) \vu{y}\right)$, 
$\omega_{\mathcal{D}} = \frac{\sqrt{3}\gamma \mathcal{D}} {M_s}$, 
$\boldsymbol{\omega}_{K, n2} = \frac{\omega_K}{4}\left(\left(n_{1, y} + n_{2, x} \right . \right.\\
\left. \left.+ \sqrt{2}m_y \right) \vu{x} + \qty(n_{1, x} + 3n_{2, y} +\sqrt{2} m_x) \vu{y}\right)$, \\
$\boldsymbol{\omega}^{ii}_{D, ni} = \frac{2 \gamma D^{ii}}{M_s}\qty(\qty(\grad \vdot \vb{n}_i)\vb{z} - \grad n_{i, z})$, \\ 
and $\boldsymbol{\omega}^{ij}_{D, ni} = \frac{\sqrt{3} \gamma D^{ij}}{M_s}\qty(\qty(\grad \vdot \vb{n}_i)\vb{z} - \grad n_{i, z})$.

The equations of motion (Eqs.~(\ref{eq:m2}) and~(\ref{eq:motion1})) derived here are
useful in the numerical study of
textures like domain walls, skyrmions, and spin-waves in AFMs with biaxial anisotropy under the effect of external magnetic field and spin current. However, here we are interested in analytically studying oscillatory dynamics of the order parameter in thin-film AFMs, therefore, we neglect inhomogeneous interactions compared to the homogeneous fields.
Using Eq.~(\ref{eq:n1_n2}) in Eq.~(\ref{eq:motion1}) and neglecting the time derivative of $n_{1z}$ and $n_{2z}$, we have
\begin{equation}\label{eq:pend1}
    \ddot{\varphi} + \alpha \omega_{E} \dot{\varphi} + \omega_{E} \frac{\omega_{K}}{2} \sin{2\varphi} + \omega_{E} \omega_{s} = 0, 
\end{equation}
where $\omega_{K} = \textcolor{black}{2}\gamma K_{e}/M_{s}$. This indicates that in the limit of strong exchange interaction, the dynamics of the staggered order parameters is identical to that of a damped-driven non-linear pendulum~\citep{coullet2005damped}. This equation is identical to the case of collinear AFMs such as NiO when the direction of spin polarization is \textcolor{black}{perpendicular to the easy-plane}~\citep{cheng2015ultrafast, khymyn2017antiferromagnetic, lee2019antiferromagnetic}. 
However, the dynamics of the non-collinear coplanar AFMs discussed here is significantly different in the direction of the spin torques, magnitude of threshold currents as well as the range of possible frequencies.
\textcolor{black}{Here, the $\sin{2 \varphi}$ dependence signifies a two-fold anisotropy symmetric system.} 

\subsection{Negative  Chirality}\label{minus}
For the case of $-\pi/2$ chirality, it can be shown that $\vb{m}$ is a dependent variable of the N\'eel order; however, in this case there are additional in-plane terms that arise due to a competition between the DMI, exchange coupling and magnetocrystalline anisotropy. To a first order, $\vb{m}$ is expressed as~\citep{liu2017anomalous, yamane2019dynamics} (also see supplementary material~\citep{supple}) 
\begin{equation}\label{eq:m3}
    \begin{split}
        \vb{m} &= - \frac{1}{\omega_{E}}\left(\vb{n}_{1} \cp \vb{\dot{n}}_{1} + \vb{n}_{2} \cp \vb{\dot{n}}_{2} + \vb{n}_{3} \cp \vb{\dot{n}}_{3} - \gamma \mu_{0} \vb{H}_{a} \right. \\
           &\left. \textcolor{black}{- \beta \omega_{s} \vb{n}_{p}}\right) - \frac{\omega_{K}}{2\omega_{E}} \qty(\cos{\varphi} \vb{x} - \sin{\varphi} \vb{y})  ,        
    \end{split}
\end{equation}
which is used to arrive at the equation of motion for the N\'eel vectors as
\begin{equation}\label{eq:motion2}
    \begin{split}
        &\vb{n}_{1} \cp \left[\vb{\ddot{n}}_{1} - c^{2} \grad^{2}{\vb{n}_{1}} - \omega_{E} \boldsymbol{\omega}_{Kn1} + \omega_{E} {\omega}_{K_{h}} \qty(\vu{n}_{1} \vdot \vb{z}) \vb{z} \right. \\
        &+ \left. \omega_{E} \omega_{\mathcal{D}}(\vb{n}_{2} \cp \vb{z}) + \omega_{E}\boldsymbol{\omega}_{Dn1}^{ii} + \omega_{E}\boldsymbol{\omega}_{Dn2}^{ij} + \alpha \omega_{E} \dot{\vb{n}}_{1} \right.\\
        &-\left. \omega_{E}\omega_{s}\qty(\vb{n}_{1} \cp \vb{n}_{p})\right] + \vb{n}_{2} \cp \left[\vb{\ddot{n}}_{2} - c^{2} \grad^{2}{\vb{n}_{2}} - \omega_{E} \boldsymbol{\omega}_{Kn2}\right. \\
        &+ \left. \omega_{E} {\omega}_{K_{h}}\qty(\vu{n}_{2} \vdot \vb{z}) \vb{z} - \omega_{E} \omega_{\mathcal{D}}(\vb{n}_{1} \cp \vb{z}) + \omega_{E}\boldsymbol{\omega}_{Dn2}^{ii} \right. \\
        &- \left. \omega_{E}\boldsymbol{\omega}_{Dn1}^{ij} + \alpha \omega_{E} \dot{\vb{n}}_{2} - \omega_{E}\omega_{s}\qty(\vb{n}_{2} \cp \vb{n}_{p})\right] + \vb{n}_{3} \cp \ddot{\vb{n}}_{3}\\
        &+ \gamma \mu_{0} \left(\vb{n}_{1} \cp \dot{\vb{n}}_{1} + \vb{n}_{2} \cp \dot{\vb{n}}_{2} + \vb{n}_{3} \cp \dot{\vb{n}}_{3}\right) \cp \vb{H}_{a} - \gamma \mu_{0} \dot{\vb{H}}_{a} \\
        &- \frac{\omega_{K}}{2} \qty(\sin{\varphi} \vb{x} + \cos{\varphi} \vb{y}) \dot{\varphi}- \gamma \mu_{0} \frac{\omega_{K}}{2}\left(H_{a, z} \sin{\varphi} \vb{x} \right. \\
        & \left. + H_{a, z} \cos{\varphi} \vb{y}  - \qty(H_{a, x} \sin{\varphi} + H_{a, y} \cos{\varphi})\vb{z}\right) = 0.
    \end{split}
\end{equation}
Similar to the previous case, we are interested in a theoretical analysis of the oscillation dynamics in thin film AFMs with negative chirality. Therefore, we use Eq.~(\ref{eq:n1_n2}) in Eq.~(\ref{eq:motion2}) and neglect all the inhomegeneous interactions to arrive at a damped-driven linear pendulum equation given as  
\begin{equation}\label{eq:pend2}
    \ddot{\varphi} + \alpha \omega_{E} \dot{\varphi} + \omega_{E} \omega_{s} = 0. 
\end{equation}
\textcolor{black}{Here the dependence of the dynamics on anisotropy is not zero but very small, and it scales proportional to $\frac{\omega_{K}^3}{\omega_E^2}\cos{6 \varphi}$~\citep{liu2017anomalous}. However, for a first-order approximation in $\vb{m}$ and dynamics in the THz regime, it can be safely ignored. The $\cos{6 \varphi}$ dependence implies that these materials host a six-fold anisotropic symmetry.}
Though this equation is similar to that obtained for the case of a collinear AFM with spin polarization along the easy axis~\citep{lee2019antiferromagnetic}, the dynamics is significantly different from
that of the collinear AFM. 

\subsection{Comparison of Dynamics for Positive and Negative Chiralities}\label{comparison}
Here, we contrast the dynamics of AFM order parameter for positive and negative chiralities.
The numerical results presented in this section are obtained in the single-domain limit assuming
thickness $d_{a} = 4 \ \mathrm{nm}$, $\alpha = 0.01$, $M_{s} = 1.63 \ \mathrm{T}$, $ \mathcal{K}_{e} = 3 \ \mathrm{MJ/m^{3}}$, $\mathcal{J} = 2.4 \times 10^{8} \ \mathrm{J/m^{3}}$, $\mathcal{D} = -20 \ \mathrm{MJ/m^{3}}$ for positive chirality or $20 \ \mathrm{MJ/m^{3}}$ for negative chirality~\citep{yamane2019dynamics}, unless specified otherwise.

Figure~\ref{fig:stationary} shows the stationary solutions of the thin-film AFM system with different chiralities.
For the case of positive chirality, it can be observed from Fig.~\ref{fig:stationary}(a) that in equilibrium the sublattice vectors $\vb{m}_{i}$ coincide with the easy axes $\vb{u}_{e, i}$.
When a non-zero spin current is applied, the equilibrium state is disturbed; however, below a certain threshold, $J_{s}^{\mathrm{th1}}$, the system dynamics converge to a stationary solution in the easy-plane of the AFM, indicated by dashed blue,  and dotted red set of arrows. An equivalent representation of the stationary solutions in terms of the staggered order parameters is presented in Fig.~\ref{fig:stationary}(b). $\vb{n}_{1}$ and $\vb{n}_{2}$ are perpendicular to each other \textcolor{black}{with zero out-of-plane component for all values of the input currents}. The gold dash-dotted arrows passing through $\vb{n}_{1}$ correspond to the stationary solutions given as $\varphi = -\frac{1}{2}\sin^{-1}\left({\frac{2\omega_{s}}{\omega_{K}}}\right)$, obtained analytically by setting both $\dot{\varphi}$ and $\ddot{\varphi}$ as zero in Eq.~(\ref{eq:pend1}). \textcolor{black}{In positive chirality material, the average magnetization $\vb{m}$ is vanishingly small in the stationary state. This can also be perceived from Eq.~(\ref{eq:m2}) as we do not consider any external field.
Since these materials have a two-fold symmetry, they also host $\varphi = \pi -\frac{1}{2}\sin^{-1}\left({\frac{2\omega_{s}}{\omega_{K}}}\right)$ stationary states.}

For the case of negative chirality, it can be observed from Fig.~\ref{fig:stationary}(c) that in equilibrium only the sublattice vector $\vb{m}_{3}$ coincides with the its corresponding easy axis, whereas both $\vb{m}_{1}$ and $\vb{m}_{2}$ are oriented such that the energy due to DMI is dominant over anisotropy, which in turn
lowers the overall energy of the system. It can be observed from Fig.~\ref{fig:stationary}(d) that $\vb{n}_{1}$ and $\vb{n}_{2}$ are almost perpendicular to each other. A small in-plane net magnetization exists in this case and is shown here as a zoomed in value (zoom factor = 100x) for the sake of comparison \textcolor{black}{to staggered order parameters}. \textcolor{black}{Due to the six-fold anisotropy dependence other equilibrium states, wherein either of $\vb{m}_{1}$ or $\vb{m}_2$ coincide with their easy axis while the other two sublattice vectors do not, also exist. Finally, due to the small anisotropy dependence, non-equilibrium stationary states exist for much lower currents~\citep{takeuchi2021chiral} than those considered here, and therefore are not shown.}

\begin{figure*}[ht!]
  \centering
  \includegraphics[width = 2\columnwidth, clip = true, trim = 0mm 2mm 0mm 0mm]{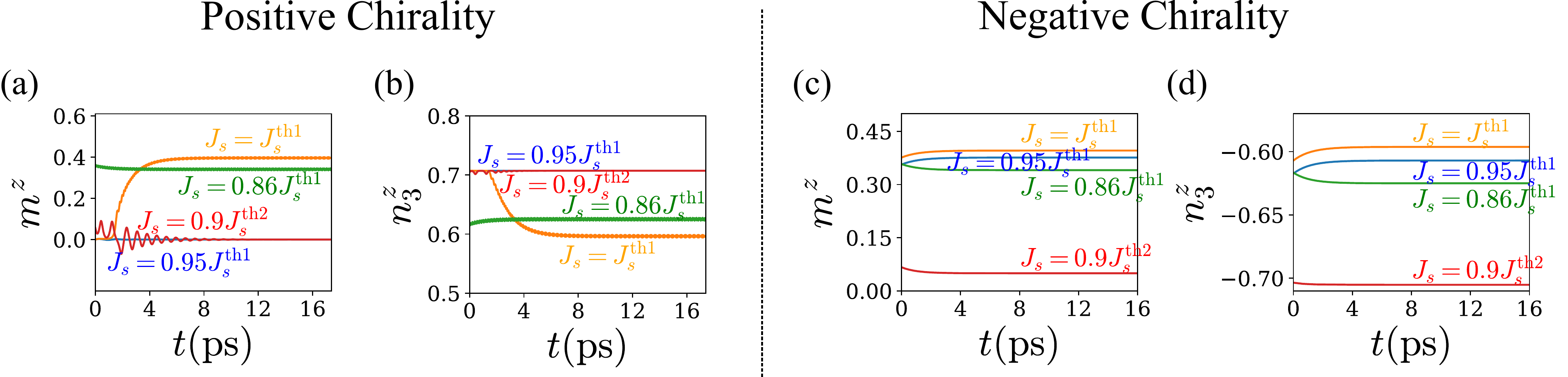}
  \caption{The out-of-plane (z) component of the average magnetization, $\vb{m}$, and $\vb{n}_{3}$ for different values of input currents for both positive and negative chirality. Positive chirality: (a) When current is increased from zero but to a value below the threshold ($0.95 J_{s}^{\mathrm{th1}}$), $m^{z}$ is zero. However, it increases to a large value when $J_{s} = J_{s}^{\mathrm{th1}}$. $m^{z}$ decreases again to a smaller value when the current is decreased to $J_{s} = 0.86J_{s}^{\mathrm{th1}}$. Finally, when the current is further reduced below the lower threshold to $0.9J_{s}^{\mathrm{th2}}$, $m^{z}$ becomes zero again. (b) $\vb{n}_{3}$ is initially equal to $1/\sqrt{2}$, but decreases in magnitude during the AFM dynamics since \textcolor{black}{the magnitude of $\vb{m}$ increases when the sublattice vectors move out of the plane}. As soon as the current is lowered below $J_{s}^{\mathrm{th2}}$, the system goes to a stationary state and $\vb{n}_{3} = 1/\sqrt{2}$. Negative chirality: (c) Since the threshold current in this case is small, non-zero $m^{z}$ is observed for all values of current considered here. (d) $\vb{n}_{3}$ decreases in magnitude when current increases but approaches $-1/\sqrt{2}$ for lower values of current. Here $\alpha = 0.01$, and $d_{a} = 4 \ \mathrm{nm}$.} 
  \label{fig:dy}
\end{figure*}

For materials with positive chirality, the system becomes unstable when
the input spin current exceeds the threshold, $J_{s}^{\mathrm{th1}}$.
The resultant spin torque pushes the N\'eel vectors out of the easy-plane, and they oscillate around the spin polarization axis, $\vb{n}_{p} = \vb{z},$ with THz frequency due to strong exchange. This threshold current is given as~\citep{khymyn2017antiferromagnetic, lee2019antiferromagnetic, parthasarathy2021precessional} 
\begin{equation}\label{eq:J1}
    J_{s}^{\mathrm{th1}} = d_{a}\frac{2e}{\hbar}\frac{M_{s}}{\gamma}  \frac{\omega_{K}}{2} = d_{a} \frac{2e}{\hbar} {\mathcal{K}_{e}},
\end{equation}
while the frequency of oscillation in the limit of large input current (neglecting the $\sin{2\varphi}$ term) from Eq.~(\ref{eq:pend1}) is given as\citep{lee2019antiferromagnetic, parthasarathy2021precessional}
\begin{equation}\label{eq:frq}
    f = \frac{1}{2\pi} \frac{\omega_{s}}{\alpha} = \frac{1}{2\pi} \frac{\hbar}{2e}\frac{\gamma J_{s}}{M_{s} d_{a}} \frac{1}{\alpha}.
\end{equation}
Additionally, for small damping,
there exists a lower threshold, $J_{s}^{\mathrm{th2}} < J_{s}^{\mathrm{th1}}$,
which is equal to the current that pumps in the same amount of energy that is lost in one time period due to damping~\citep{khymyn2017antiferromagnetic, lee2019antiferromagnetic, parthasarathy2021precessional}. The lower threshold current
is given as
\begin{equation}\label{eq:J2}
    J_{s}^{\mathrm{th2}} = d_{a}\frac{2e}{\hbar}\frac{M_{s}}{\gamma}  \frac{2 \alpha}{\pi}\sqrt{\omega_{E}\omega_{K}} = d_{a}\frac{2e}{\hbar} \frac{2 \alpha}{\pi}\sqrt{\textcolor{black}{6}\mathcal{J}\mathcal{K}_{e}}.
\end{equation}
The presence of two threshold currents enables energy-efficient operation of the THz oscillator in the hysteretic region~\citep{khymyn2017antiferromagnetic}. 

The average frequency response as a function of the input spin current and the Fourier transform of the oscillation dynamics is plotted in the left panel of Fig.~\ref{fig:freq} for materials with positive chirality. 
It can be observed from Fig.~\ref{fig:freq}(a) that the fundamental frequency for different film thicknesses scales as predicted by Eq.~(\ref{eq:frq}) except for low currents near $J_{s}^{\mathrm{th2}}$ where non-linearity in the form of higher harmonics appears as seen from the FFT response in Fig.~\ref{fig:freq}(e). Next, Figs.~\ref{fig:freq}(b), (f) show that the non-linearity in the frequency response for low input current increases as the value of the damping coefficient increases. This is expected as \textcolor{black}{the contribution from the uniaxial anisotropy ($\sin{2\varphi}$ term) becomes significant owing to large damping and low current} making the motion non-uniform ($\ddot{\varphi} \neq 0$)~\citep{khymyn2017antiferromagnetic}. 

In the case of materials with negative chirality, \textcolor{black}{small equivalent anisotropy} suggests that the threshold current for the onset of oscillations is \textcolor{black}{very small}, while the frequency of oscillations increases linearly with the spin current \textcolor{black}{considered here} and is given by
Eq.~(\ref{eq:frq}). Indeed the same can be observed from Figs.~\ref{fig:freq}(c), (d) where the results of numerical simulations exactly match 
the 
analytic expression. The FFT signal
in Figs.~\ref{fig:freq}(g), (h) contains only one frequency corresponding to uniform rotation of the order
($\ddot{\varphi} = 0$).
This coherent rotation of the order parameter with \textcolor{black}{such small threshold current~\citep{takeuchi2021chiral}} in AFM materials with negative chirality opens up the
possibility of operating such AFM oscillators at very low energy for frequencies ranging from MHz-THz. 
It can also be observed from Fig.~\ref{fig:freq}(b), (d) that for lower values of damping, such as $\alpha = 0.005$, the frequency of oscillations saturates for input current slightly above $J_{s}^{\mathrm{th1}}$. This is because the energy pumped into the system is larger than that dissipated by damping. As a result the sublattice vectors move out of the easy-plane and get oriented along the spin polarization direction (slip-flop). \textcolor{black}{For larger values of damping, the same would be observed for larger values of current.}
Finally, we would like to point out that the values of both $J_{s}^{\mathrm{th1}}$ and $J_{s}^{\mathrm{th2}}$ observed from numerical simulations were slightly different from their analytical values for different damping constants, similar to that reported in Ref.~\citep{puliafito2019micromagnetic} for collinear AFMs. 

Figure~\ref{fig:dy} shows the out-of-plane (z) components of $\vb{m}$ and $\vb{n}_{3}$ for non-zero currents
for AFMs with different chiralities. For negative chirality materials, the steady-state $z$ component of both $\vb{m}$ and $\vb{n}_{3}$ does not oscillate with time ($\ddot{\varphi} \approx 0$), whereas, for the case of positive chirality, the steady-state $z$ components of both $\vb{m}$ and $\vb{n}_{3}$ show small oscillations with time ($\ddot{\varphi} \neq 0$) similar to the case of $\mathrm{NiO}$ with spin polarization along the hard axis~\citep{khymyn2017antiferromagnetic}. 

It can be observed from Fig.~\ref{fig:dy}(a) that for positive chirality, as current increases from below the upper threshold current (0.95$J_{s}^{\mathrm{th1}}$) to $J_{s}^{\mathrm{th1}}$, the out-of-plane component of magnetization vectors and hence the average magnetization $\vb{m}$ increases from zero to a larger value. Due to the hysteretic nature of the AFM oscillator, the magnitude of $\vb{m}$ reduces when current is lowered but is non-zero as long as the input current is above $J_{s}^{\mathrm{th2}}$. Similarly, it can be observed from Fig.~\ref{fig:dy}(b) that the out-of-plane component of $\vb{n}_{3}$, which was initially $1/\sqrt{2}$, decreases as the current increases above $J_s^\mathrm{th1}$.
When the current is lowered to a value below $J_{s}^{\mathrm{th1}}$ ($0.86J_{s}^{\mathrm{th1}}$ here) ,
the magnitude of the out-of-plane component of $\vb{n}_{3}$ increases again and eventually
saturates to $1/\sqrt{2}$ when the current is lowered further below $J_{s}^{\mathrm{th2}}$ ($0.9J_{s}^{\mathrm{th2}}$, here).

It can also be observed from Fig.~\ref{fig:dy}(c) that for negative chirality AFMs, the out-of-plane component of the average magnetization although small is non-zero even for small currents due to \textcolor{black}{the lower value of} threshold current.
On the other hand, $n_{3}^{z}$ in Fig.~\ref{fig:dy}(d) decreases in magnitude \textcolor{black}{from an initial value of $\lambda = 1/\sqrt{2}$ to $\lambda < 1/\sqrt{2}$} as current increases since $\norm{\vb{m}}$ increases.
The values of current are assumed to be the same for both positive and negative chirality AFMs for the sake of comparison.
\begin{figure}[ht!]
  \centering
  \includegraphics[width = \columnwidth, clip = true, trim = 0mm 2mm 0mm 0mm]{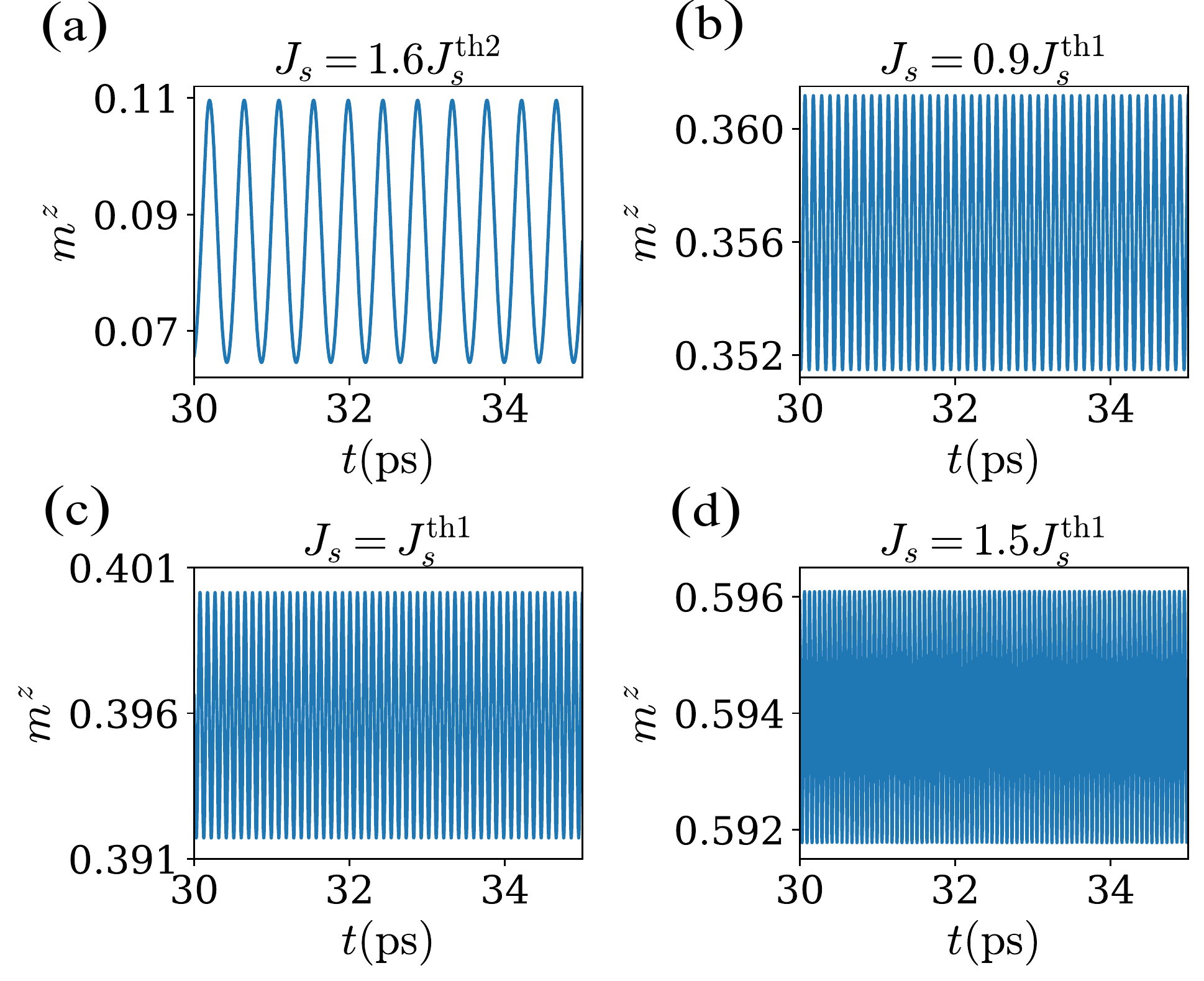}
  \caption{$m^{z}$ for four different values of input current $J_{s}$. ${m}^{z}$ increases with current and so does the frequency of oscillation. Here (a) $J_{s} = 1.6J_{s}^{\mathrm{th2}}$, (b) $J_{s} = 0.9J_{s}^{\mathrm{th1}}$. They are both inside the hysteretic region bounded by $J_{s}^{\mathrm{th2}}$ and $J_{s}^{\mathrm{th1}}$. (c) $J_{s} = J_{s}^{\mathrm{th1}}$, and (d) $J_{s} = 1.5J_{s}^{\mathrm{th1}}$ lies outside the hysteretic region. 
  These results correspond to $\alpha = 0.01$, and $d_{a} = 4 \ \mathrm{nm}$.} 
  \label{fig:mz}
\end{figure}  
Next, using Eq.~(\ref{eq:n1_n2}) in Eq.~(\ref{eq:m2}), it can be shown that $\dot{\varphi}$ is directly proportional to ${m}^{z}$~\citep{yamane2019dynamics}. Therefore, to present the features of angular velocity with input current, we show ${m_{z}}$ for four different values of input current $J_{s}$ for positive chirality material in Fig.~\ref{fig:mz} . Here, Figs.~\ref{fig:mz}(a)-(c) correspond to the hysteretic region, whereas Fig.~\ref{fig:mz}(d) is for current outside the hysteretic region.
As mentioned previously, an increase in current increases the spin torque on the sublattice vectors which leads to an increase in ${m}^{z}$ and hence $\dot{\varphi}$. 



\section{Signal Extraction}\label{extract}
An important requirement for the realization of an AFM-based auto-oscillator is the extraction of the generated THz oscillations as measurable electrical
quantities viz. voltage and current.
It is expected for the extracted voltage signal to oscillate at the same frequency as that of the N\'eel vector and contain substantial output power ($ > 1 \ \mathrm{\mathrm{\mu W}}$)~\citep{sulymenko2018terahertz}. 
In this regard, the landmark theoretical work on $\mathrm{NiO}$ based oscillator~\citep{khymyn2017antiferromagnetic} suggested the measurement of spin pumped~\citep{li2020spin} time varying inverse spin Hall voltage~\citep{hou2019spin} across the heavy metal $(\mathrm{Pt})$ of a $\mathrm{NiO/Pt}$ heterostructure.
However, the time varying voltage at THz frequency requires an AFM with significant in-plane biaxial anisotropy~\citep{khymyn2017antiferromagnetic, sulymenko2018terahertz}, thus limiting the applicability of this scheme to only select AFM materials. In addition, the output power of the generated signal is sizeable (above $1 \ \mathrm{\mathrm{\mu W}}$) only for frequencies below $0.5 \  \mathrm{THz}$~\citep{sulymenko2018terahertz}. 
A potential route to overcoming the aforementioned limitations is coupling the AFM signal generator to a high-Q dielectric resonator, which would enhance the output power 
even for frequencies above $0.5 \  \mathrm{THz}$~\citep{sulymenko2017terahertz}. This method, however, requires devices with sizes in the 10's micrometers range for frequencies above $2$ THz and for the AFMs to possess a tilted net magnetization in their ground state~\citep{sulymenko2018terahertz}. 
A more recent theoretical work on collinear AFM THz oscillators~\citep{puliafito2019micromagnetic} suggested employing Anisotropy Magnetoresistance (AMR) or Spin Magnetoresistance (SMR) measurements in a four terminal AFM/HM spin Hall heterostructure. This would enable the extraction of the THz oscillations as longitudinal or transverse voltage signals. However, the reported values of both AMR and SMR at room temperatures in most AFMs is low and would, in general, require modulating the band structure for higher values~\citep{bai2020functional}. 

\begin{figure}[ht!]
  \centering
  \includegraphics[width = \columnwidth, clip = true, trim = 0mm 0mm 0mm 0mm]{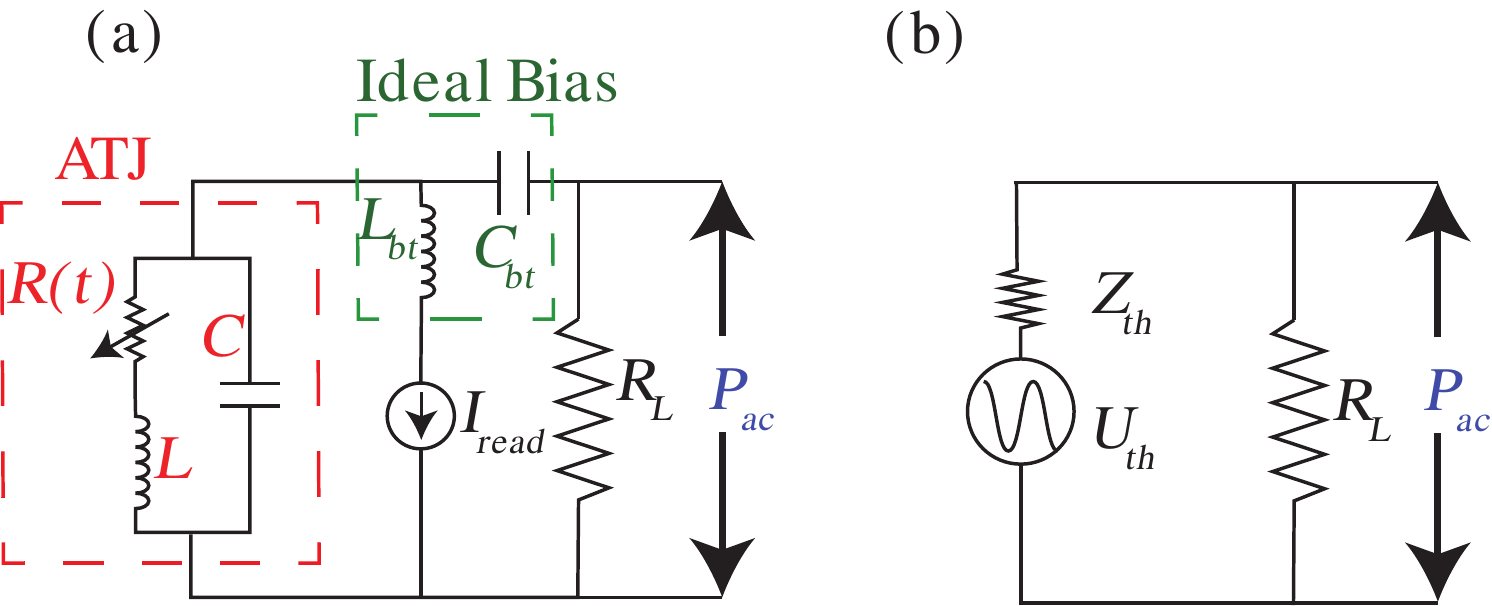}
  \caption{(a) An equivalent circuit representation of Fig.~\ref{fig:geometry} (adapted from~\citep{sulymenko2018terahertz}). The generation (write) current is not shown in the circuit, although its effect is included as a variation in the resistance $R(t)$ through its frequency dependence.
  (b) Thevenin equivalent of (a).} 
  \label{fig:ck_diag}
\end{figure} 
A recent theoretical work~\citep{sulymenko2018terahertz} proposed employing a four terminal AFM tunnel junction (ATJ) in a spin Hall bilayer structure with a conducting AFM to effectively generate and detect THz frequency oscillations as variations in the tunnel anisotropy magnetoresistance~\citep{park2011spin}. 
A DC current passed perpendicularly to the plane of the ATJ generates an AC voltage, which is measured across an externally connected load. 
It was shown that both the output power and its efficiency decrease as frequency increases, nevertheless, it was suggested that this scheme could be used for signal extraction in the frequency range of $0.1-10$ THz, although the lateral size of the tunnel barrier required for an optimal performance depends on the frequency of oscillations (size decreases as the frequency increases)~\citep{sulymenko2018terahertz}. 
The analysis presented in Ref.~\citep{sulymenko2018terahertz}, however, neglects the generation current compared to the read current while evaluating the efficiency of power extraction. 
But it can be observed from the results in Section~\ref{comparison} that the threshold current, and, therefore, the generation current, depend on AFM material properties, such as damping, anisotropy, and exchange constants, and could be quite large. Therefore, in our work we include the effect of the generation current to accurately model the power efficiency of the TAMR scheme. 
\begin{figure}[ht!]
  \centering
  \includegraphics[width = \columnwidth, clip = true, trim = 0mm 2mm 0mm 0mm]{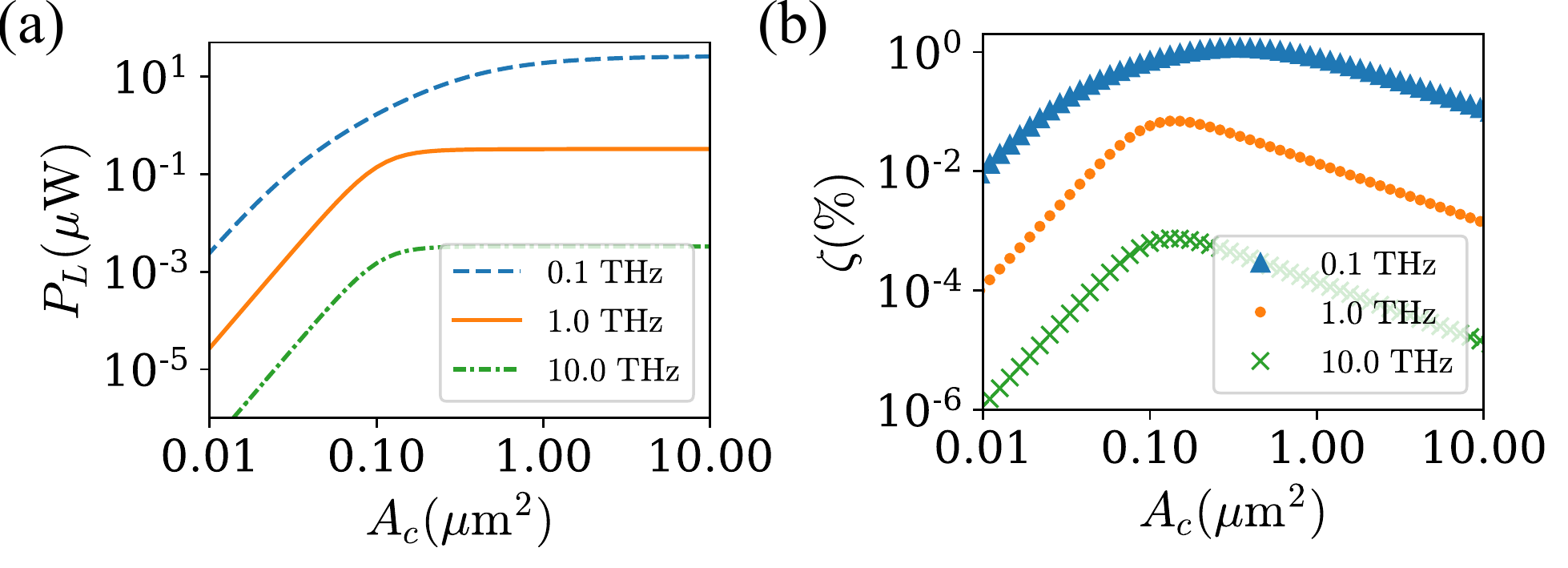}
  \caption{(a) Output power and (b) efficiency dependence on the area of cross-section of the tunnel barrier for different frequencies. The thickness of the barrier is fixed to $d_{b} = 1 \ \mathrm{nm}$. The effect of write current and the input power associated with it is not considered here, therefore, these results are independent of the choice of the AFM material.} 
  \label{fig:PL_zeta}
\end{figure}  

In order to evaluate the performance of the TAMR scheme, an equivalent circuit representation (adapted from Ref.~\citep{sulymenko2018terahertz}) of the device setup of Fig.~\ref{fig:geometry} is shown in Fig.~\ref{fig:ck_diag}(a), 
while its Thevenin equivalent representation is shown in Fig.~\ref{fig:ck_diag}(b). 
The circuits in Fig.~\ref{fig:ck_diag} only represent the read component, while the THz generation component is omitted for the sake of clarity.
In Fig.~\ref{fig:ck_diag}(a), the dashed red box encloses a circuit representation of the ATJ, comprising a series combination of an oscillating resistance 
$R(t) = R_{0} + \Delta R \cos{\omega t}$ and inductance $L = \mu_{0} d_{b}$, connected in parallel to a junction capacitor $C = A_{c} \epsilon \epsilon_{0} /d_{b}$ (assumed parallel plate). 
The constant component, $R_0$, in the oscillating resistance, $R(t)$, is the equilibrium resistance of the $\mathrm{MgO}$ barrier and is given as $R_{0} = \frac{R A(0) \exp(\kappa d_{b})}{A_{c}}$. 
Here, $R A(0)$ is the resistance-area product of a zero-thickness tunnel barrier, $\kappa$ is the tunneling parameter, $d_{b}$ is the barrier thickness, and $A_{c}$ is the cross-sectional area. 
The pre-factor, $\Delta R$, of the time varying component of $R(t)$ is the resistance variation due to the oscillation of the magnetization vectors with respect to the polarization axis. $\Delta R = \qty(\eta/(2 + \eta)) R_{0}$, where $\eta$ is the TAMR ratio of the barrier and depends on the temperature and material properties. 

\begin{table}[t!]\caption{\label{tab:materials}{List of common antiferromagnetic materials $\mathrm{Mn_{3}X}$ and their associated parameters. Here $M_{s}$ is in Tesla, $\mathcal{K}_{e}$ is in $\mathrm{kJ/m^{3}}$}, and $\mathcal{J}$ is in $\mathrm{MJ/m^{3}}$. Sign of $\mathcal{D}$ which decides the chirality is also mentioned.}  
\begin{ruledtabular}
\begin{tabular}{lcccccc}
X  & $\alpha$ & $M_{s}$  & $\mathcal{K}_{e}$ & $\mathcal{J}$ & $\mathcal{D}$ & Ref.\\
\hline
$\mathrm{Ir}$ & $0.01$ & $1.63$ & $3000$ & $240$ & - &\citep{yamane2019dynamics}\\
$\mathrm{Pt}$ & $0.013$ & $1.37$ & $10$ & $280$ & - &\citep{kren1968magnetic}\\
$\mathrm{Rh}$ &  $0.013$ & $2.00$ & $10$ & $230$ & - &\citep{kren1968magnetic, feng2015large, zhang2017strong}\\
{$\mathrm{Ga}$} &  $0.008$ & $0.54$ & $100$ & $110$ & + &\citep{kurt2011mn3, zhang2017strong, nyari2019weak, seyd2020mn3ge}\\
{$\mathrm{Sn}$} &  $ 0.003 $ & $0.50$ & $110$ & $59$ & + &\citep{tsai2020electrical, liu2017anomalous, zhang2017strong, nyari2019weak}\\
{$\mathrm{Ge}$} &  $0.0009$ & $0.28$ & $1320$ & $77$ & + &\citep{yamada1988magnetic, zhang2017strong, nyari2019weak, seyd2020mn3ge}\\
{$\mathrm{GaN}$} &  $0.1$ & $0.69$ & $10$ & $280$ & - &\citep{gurung2020spin, chen2017electronic}\\
{$\mathrm{NiN}$} &  $0.1$ & $1.54$ & $10$ & $177$ & - &\citep{gurung2020spin, chen2017electronic}\\
\end{tabular}
\end{ruledtabular}
\end{table}

Due to the flow of the DC current, $I_{\mathrm{read}}$, an alternating voltage develops across the ATJ, which is measured across an externally connected load $R_{L}$, separated from the ATJ via an ideal bias tee (enclosed in the green dashed box). The bias-tee, characterized by an inductance $L_{bt}$ and a capacitance $C_{bt}$, and assumed to have no voltage drop across it, blocks any DC current from flowing into the external load. Therefore, the AC voltage of the ATJ is divided only into its impedance (a combination of $R_{0}$, $L$, and $C$) and that of the load $R_{L}$~\citep{sulymenko2018terahertz}. 

Next, we simplify the ATJ circuit into a Thevenin impedance $Z_{th}$ and voltage $U_{th}$ as shown in Fig.~\ref{fig:ck_diag}(b). They are evaluated as
\begin{equation}\label{eq:zth}
    Z_{th} = \frac{R_{0} + j \omega L}{(1 - \xi) + j \beta},
\end{equation}
and 
\begin{equation}\label{eq:uth}
    U_{th} = \frac{U_{ac}}{(1 - \xi) + j \beta},
\end{equation}
where $j = \sqrt{-1}$, $\omega = 2 \pi f$, $\xi = \omega^{2} L C$, $\beta = \omega R_{0} C$, and $U_{ac} = I_{\mathrm{read}} \Delta R$. The output voltage and average power across the load can then be obtained as
\begin{equation}\label{eq:UL}
    U_{L} = U_{th}\frac{R_{L}}{Z_{th} + R_{L}} = U_{ac}\frac{r}{1 + j p + r(1 - \xi + j \beta)},
\end{equation}
and 
\begin{equation}\label{eq:PL}
    P_{L} = \frac{1}{2}\frac{|U_{L}|^{2}}{R_{L}} = \frac{U_{ac}^{2}}{2 R_{L}}\frac{r^{2}}{1 + q r^{2} + 2r + p^{2}},
\end{equation}
where $r = R_{L}/R_{0}$, $q = (1 - \xi)^{2} + \beta^{2}$, and $p = \frac{\omega L}{R_{0}}$. Finally, the efficiency of the power extraction can be obtained as
\begin{align}\label{eq:eff}
    \begin{split}
        \zeta &= \frac{P_{L}}{P_{in}} \\
              &= \frac{0.5 r}{1 + q r^{2} + 2r + p^{2}} \frac{1}{I_{\mathrm{write}}^{2}R_\mathrm{Gen}R_{0}/U_{ac}^{2} + 1},
    \end{split}
\end{align}
where $R_\mathrm{Gen}$ is the resistance faced by the generation current. It can be observed from Eqs.~(\ref{eq:UL})-(\ref{eq:eff}) that the output voltage, output power, and the efficiency of power extraction decrease with an increase in frequency since $\xi$, $\beta$, $q$, and $p$ increase with $\omega$~\citep{sulymenko2018terahertz, artemchuk2020terahertz}. 

Considering that the load impedance is fixed to $50 \ \Omega$ by the external circuit, one can only optimize the source impedance to achieve $P_{L} > 1 \ \mathrm{\mu W}$ and $U_{L} > 1 \ \mathrm{mV}$.
\begin{table}[t!]\caption{\label{parameters} Material Parameters of the NM, and at the NM/AFM interface.} 
\begin{ruledtabular}
\begin{tabular}{ccc}
Parameters  & Values & Ref.\\
\hline
$g_{M}$ & $3.8 \times 10^{10}$ S/m$^{2}$ & \citep{skarsvaag2015spin}\\
$g_{m}$ & $3.8 \times 10^{9}$ S/m$^{2}$ & \citep{skarsvaag2015spin}\\
$\rho_{Cu}$ & $6 \times 10^{-9} \Omega \ \mathrm{m^{2}}$ & \citep{skarsvaag2015spin}\\
$t_{Cu}$ & $5 \ \mathrm{nm}$ & \citep{skarsvaag2015spin}\\
\end{tabular}
\end{ruledtabular}
\end{table}
In this regard, the resistance of the source tunnel barrier can be altered by either
varying the thickness of the tunnel barrier, $d_{b}$, or its cross-sectional area, $A_{c}$. However, the optimum values of $d_{b}$ and $A_{c}$ for the desired output signals is frequency dependent, and, therefore, tunnel barriers of different sizes would be required for different operating frequencies~\citep{sulymenko2018terahertz, artemchuk2020terahertz}. 
For all estimates, we consider $d_{b} = 1 \ \mathrm{nm}$, $\eta = 1.3$, $\kappa = 5.6 \ \mathrm{nm^{-1}}$, $RA(0) = 0.14 \ \mathrm{\Omega \ \mu m^{2}}$, and $\epsilon = 9.8$~\citep{sulymenko2018terahertz}. 
For reliable operation of the tunnel barrier, we consider the electric field across the barrier to be $E = 0.3 \  \mathrm{V/nm}$~\citep{sulymenko2018terahertz}, which is below the barrier breakdown field.
Ignoring the effect of the generation current in Eq.~(\ref{eq:eff}), as suggested in Ref.~\citep{sulymenko2018terahertz}, we deduce from Fig.~\ref{fig:PL_zeta} 
that the optimal cross-sectional area $A_c \approx 0.36$ $\mu$m$^2$ for $f=0.1$ THz, $A_c \approx 0.25$ $\mu$m$^2$ for $f=1$ THz, $A_c \approx 0.16$ $\mu$m$^2$ for $f=10$ THz. 
\begin{figure}[ht!]
  \centering
  \includegraphics[width = \columnwidth, clip = true, trim = 0mm 0mm 0mm 0mm]{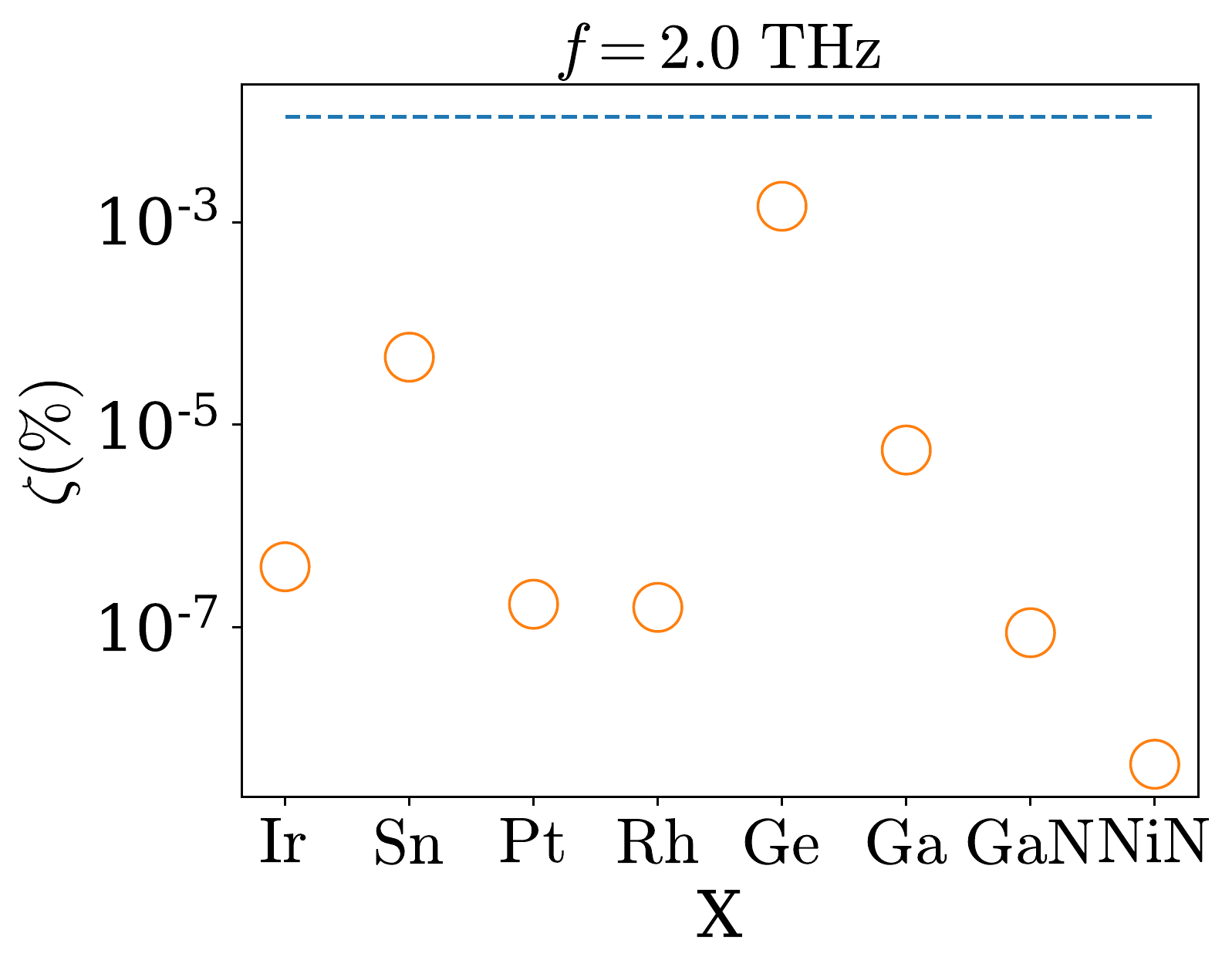}
  \caption{Power efficiency for different materials $\mathrm{Mn_{3}X}$ listed in Table~\ref{tab:materials}. The dashed horizontal line shows the expected efficiency of $\eta = 0.011 \%$ for the optimized geometry ($d_{b} = 1 \ \mathrm{nm}$, and $A_{c} = 0.24 \ \mu \mathrm{m^{2}}$) if write current is neglected. The efficiency, however, decreases significantly due to the inclusion of write current. Here the AFM thin-film thickness $d_{a}$ is assumed to be $4 \ \mathrm{nm}$. } 
  \label{fig:zeta}
\end{figure} 

Table~\ref{tab:materials} lists the material properties of various conducting AFMs.
Depending on the sign of their DMI constant, these AFMs could host moments with either a positive or a negative chirality. 
The closed-form model presented in  Eq.~(\ref{eq:frq}) can be used to evaluate the required spin current for frequency $f = 2 \ \mathrm{THz}$, regardless of the chirality since frequency scales linearly with the input current in this region (see Fig.~\ref{fig:freq}). 
For a given spin current density ($J_s$), the charge current density ($J_\mathrm{write}$) for the lateral spin-valve structure of Fig.~\ref{fig:geometry}(a) is given as
\begin{equation}
    J_{\mathrm{write}} = \frac{g_{M} + g_{m}}{g_{M} - g_{m}} J_{s}. 
\end{equation}
where $g_{M}$ and $g_m$ are the conductance of the majority- and minority-spin electrons at the NM (Cu)/FM interface.
The input power required to start the oscillations is given as $(J_{\mathrm{write}} A_{c})^{2} R_{Cu}$,
where $R_{\mathrm{Gen}} = R_{Cu} = \rho_{Cu} \frac{L_{Cu}}{A_{Cu}} = \rho_{Cu} \frac{\sqrt{A_{c}}}{t_{Cu} \sqrt{A_{c}}}$ is the resistance of the copper (NM) underneath the bottom MgO. In order to evaluate the resistance of the copper layer, we have assumed its length and width to be the same as MgO and the AFM thin-film.

The efficiency of power extraction for the listed AFM materials is presented in Fig.~\ref{fig:zeta}. The dashed horizontal line denotes the expected efficiency of $\eta = 0.011 \%$ if the effect of generation current is neglected and the area of cross-section of MgO is optimized for $f = 2.0$ THz. However, it can be observed that the efficiency decreases significantly i.e. by a few orders when the input power due to the generation current in included in the analysis. For materials with large damping and large uniaxial anisotropy constants, the required generation current is higher leading to lower efficiency. 
This result shows that further optimization of the device geometry for different materials is required to increase the efficiency. 

\begin{figure}[ht!]
  \centering
  \includegraphics[width = \columnwidth, clip = true, trim = 0mm 2mm 0mm 0mm]{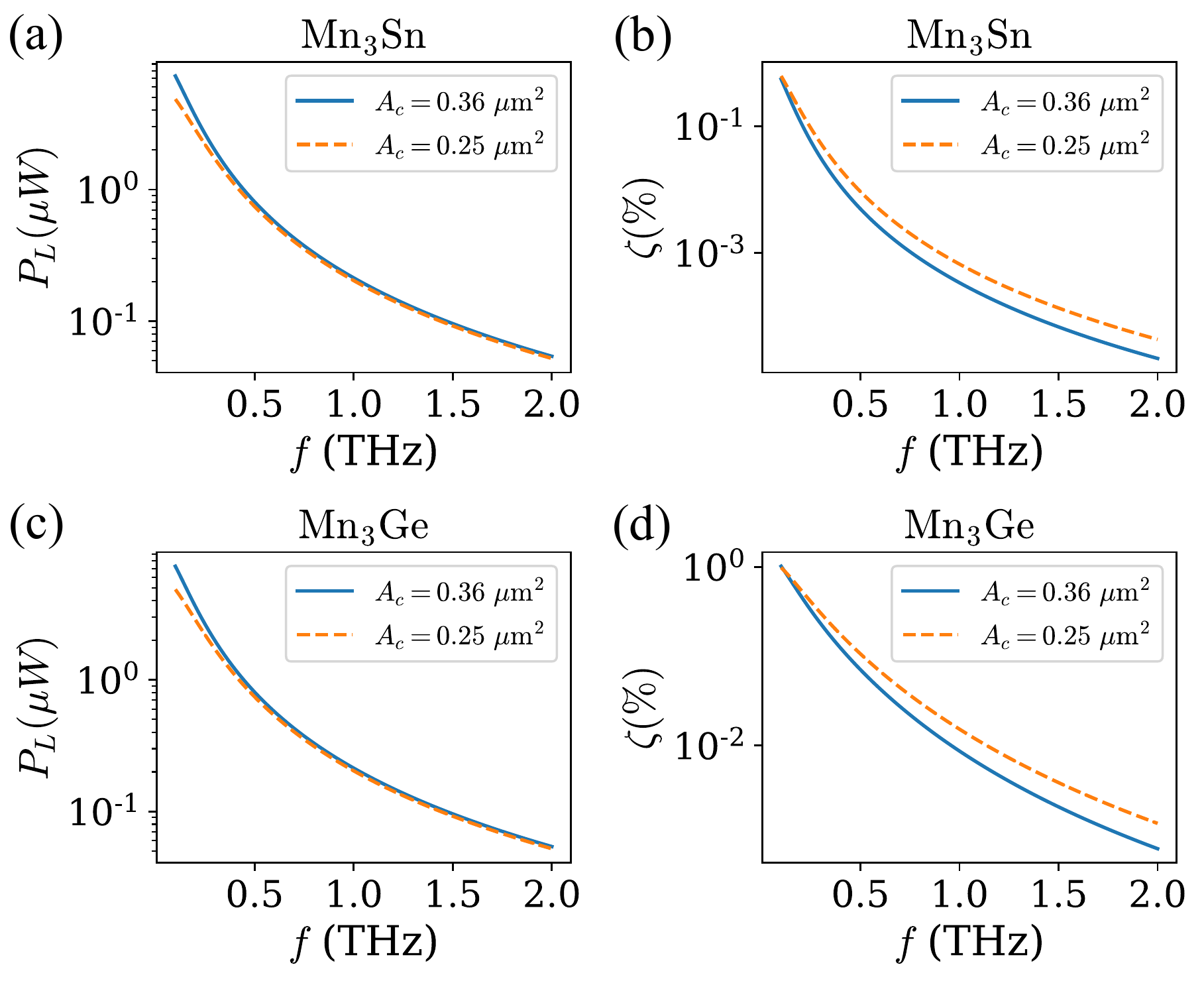}
  \caption{Upper panel: (a) Output power and (b) efficiency for Mn$_3$Sn. Lower panel: (c) Output power and (d) efficiency for Mn$_3$Ge. Two different cross-section size of the MgO barrier is considered. The output power depends only on the frequency of oscillation and therefore is same for both the materials. The efficiency of power extraction depends on the generation current, which is lower for Mn$_3$Ge, leading to a higher value of efficiency in that case.} 
  \label{fig:PL_zeta2}
\end{figure}

This method of power extraction could be more suitable for materials with negative chirality. We can observe from Fig.~\ref{fig:PL_zeta2} that the output power as well as the efficiency for both Mn$_3$Sn and Mn$_3$Ge for frequencies between 0.1 THz and 2.0 THz are significant. The required generation current for Mn$_3$Ge is smaller than that for Mn$_3$Sn, therefore, the efficiency is higher for the former. Also, the efficiency of power extraction increases with decrease in area of cross-section in both the cases but this is accompanied by a decrease in output power.

It might be possible to increase the output power and overall efficiency of the system if the material properties of the tunnel barrier such as $\eta, \kappa$, and $RA(0)$ could be altered. Large room temperature tunneling magnetoresistance in an ATJ is feasible either by using a tunnel barrier other than $\mathrm{MgO}$~\citep{su2019large} or inserting a diffusion barrier to enhance magneto-transport~\citep{zhang2018enhancement}.
Here we adopted the TAMR extraction scheme because we have considered metallic AFMs so a DC current through the ATJ structure can be easily applied. In addition, the three- or four-terminal compact ATJ structure along with its small lateral size enables dense packing of several such THz oscillators on a chip 
accompanied with a net increase in the output power and efficiency of the oscillator array~\citep{bai2020functional}. For example, with an array of $10\times 10$ such AFM oscillators excited in parallel, the output power and efficiency could be scaled up by 100$\times$ compared to the results presented in Figs.~\ref{fig:PL_zeta} and \ref{fig:zeta}.



\section{Effects of inhomogeneity due to Exchange Interaction}\label{micro}
The results presented in Section~\ref{comparison} correspond to the case of a single-domain AFM particle and are, therefore, independent of the lateral dimensions of the thin-film. This can also be deduced from the equations of the threshold current and the average oscillation frequency. 
However, when the lateral dimensions of the AFM thin-film exceed several 10's of nm, micromagnetic analysis must be carried out. In this section, we analyze the dynamics in thin-film AFMs of varying dimensions within a micromagnetic simulation framework.
\begin{figure}[ht!]
  \centering
  \includegraphics[width = \columnwidth, clip = true, trim = 0mm 0mm 0mm 0mm]{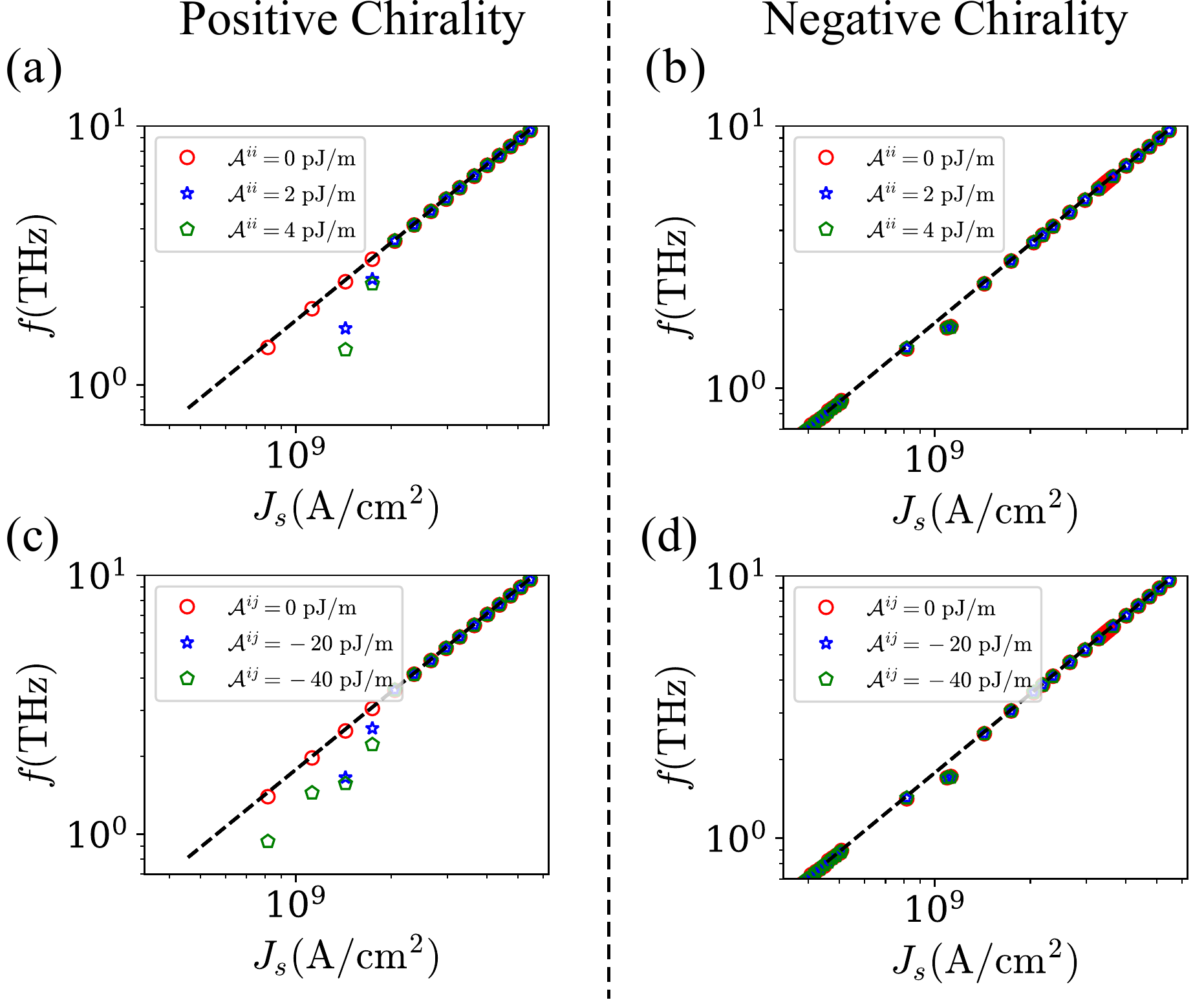}
  \caption{Frequency vs. input current for different values of inhomogeneous exchange constants (intra-sublattice (a, c), inter-sublattice (b, d)) for both positive and negative chirality. In all cases $\alpha = 0.01$, and $d_{a} = 4 \ \mathrm{nm}$. Other parameters correspond to those of $\mathrm{Mn_{3}Ir}$ as listed in Table~\ref{tab:materials} \textcolor{black}{for both positive and negative chirality materials with the exception of the sign of $\mathcal{D}$ for the latter}.} 
  \label{fig:inhomo}
\end{figure} 
We consider AFM thin-films of dimensions $50 \ \mathrm{nm} \times 50 \ \mathrm{nm}$ and investigate the effect of the inhomogeneity due to exchange interactions. In each case, the thin-film was divided into smaller cubes, each of size $1 \ \mathrm{nm} \times 1 \ \mathrm{nm} \times d_{a} \ \mathrm{nm}$, since the domain wall width $\Delta_{0} = \sqrt{(2\mathcal{A}^{ii} - \mathcal{A}^{ij})/(2\mathcal{K}_{e})} > 1 \ \mathrm{nm}$ for $\mathcal{K}_{e}$ corresponding to $\mathrm{Mn_{3}Ir}$ as listed in Table~\ref{tab:materials}. It can be observed from Fig.~\ref{fig:inhomo} that for materials with positive chirality the effects of inhomogeneity becomes important for low currents. On the other hand, for materials with negative chirality, inhomogeneities do not appear to have any effect. For positive chirality materials, the numerical values of frequency for different spring constants deviates significantly from that obtained from the single domain solution, as well as analytic results. In this case, the hysteretic region reduces in size since the lower threshold current increases in magnitude as compared to the theoretical prediction as can be observed from Fig.~\ref{fig:inhomo}(a). While we have not included the effect of inhomogeneous DMI in our work, we expect such interactions to lead to the formation of domain walls in the thin-film similar to the case of collinear AFMs~\citep{puliafito2019micromagnetic}. A more detailed analysis of the dynamics of the positive chirality materials due to variation in exchange interaction as well as inhomogeneous DMI would be carried out in a future publication.



\section{Discussion}\label{disc}
We focused on the dynamics of the order parameters in \textcolor{black}{exchange dominant} non-collinear coplanar AFMs with both positive ($+\pi/2$) and negative ($-\pi/2$) chiralities associated to the orientation of equilibrium magnetization vectors. 
\textcolor{black}{In both these classes of AFMs, the exchange energy is minimized for a $2\pi/3$ relative orientation between the sublattice vectors. Next, the negative (positive) sign of the iDMI coefficient minimizes the system energy for counterclockwise (clockwise) ordering of $\vb{m}_{1}, \vb{m}_{2}$, and $\vb{m}_{3}$ in the $\vb{x-y}$ plane leading to positive (negative) chirality.
Finally, all the sublattice vectors coincide with their respective easy axis only in the case of the positive chirality materials due to the relative anticlockwise orientation of the easy axes. On the other hand, the negative chirality materials have a six-fold symmetry wherein only one of the sublattice vectors can coincide with its respective easy axis.}
\textcolor{black}{As a result, these} AFM materials with different chiralities have significantly distinct dynamics in the presence of an input spin current. 
For AFM materials with $+\pi/2$ chirality, oscillatory dynamics are excited only when the injected spin current overcomes the anisotropy, thus indicating the presence of a \textcolor{black}{larger} current threshold. Moreover, the dynamics in such AFMs is hysteretic in nature.
Therefore, it is possible to sustain oscillations by lowering the current below that required to initiate the dynamics as long the energy pumped in by the current overcomes that dissipated by damping. On the other hand, in the case of $-\pi/2$ chirality AFMs, 
oscillations can be excited \textcolor{black}{when significantly smaller} spin current with appropriate spin polarization is injected into the AFM. Hence, $-\pi/2$ chirality AFMs may be more amenable to tuning the frequency response over a broad frequency range, from the MHz to the THz range~\citep{takeuchi2021chiral}.
The oscillation of the AFM N\'eel vectors can be measured as a coherent AC voltage with THz frequencies across an externally connected resistive load through the tunnel anisotropic magnetoresistance measurements for both $+\pi/2$ and $-\pi/2$ chirality materials. In general, as the frequency increases, the magnitude of both the output power and the efficiency of power extraction decrease, however, it is possible to enhance both these quantities by optimizing the cross-sectional area of the tunnel junction. This, however, is limited due to larger threshold current requirement for materials with large damping. Therefore, a hybrid scheme of electrically synchronized AFM oscillators on a chip could be used to further enhance the power and efficiency~\citep{grollier2006synchronization, georges2008impact}.
\begin{figure}[ht!]
  \centering
  \includegraphics[width = \columnwidth, clip = true, trim = 0mm 2mm 0mm 0mm]{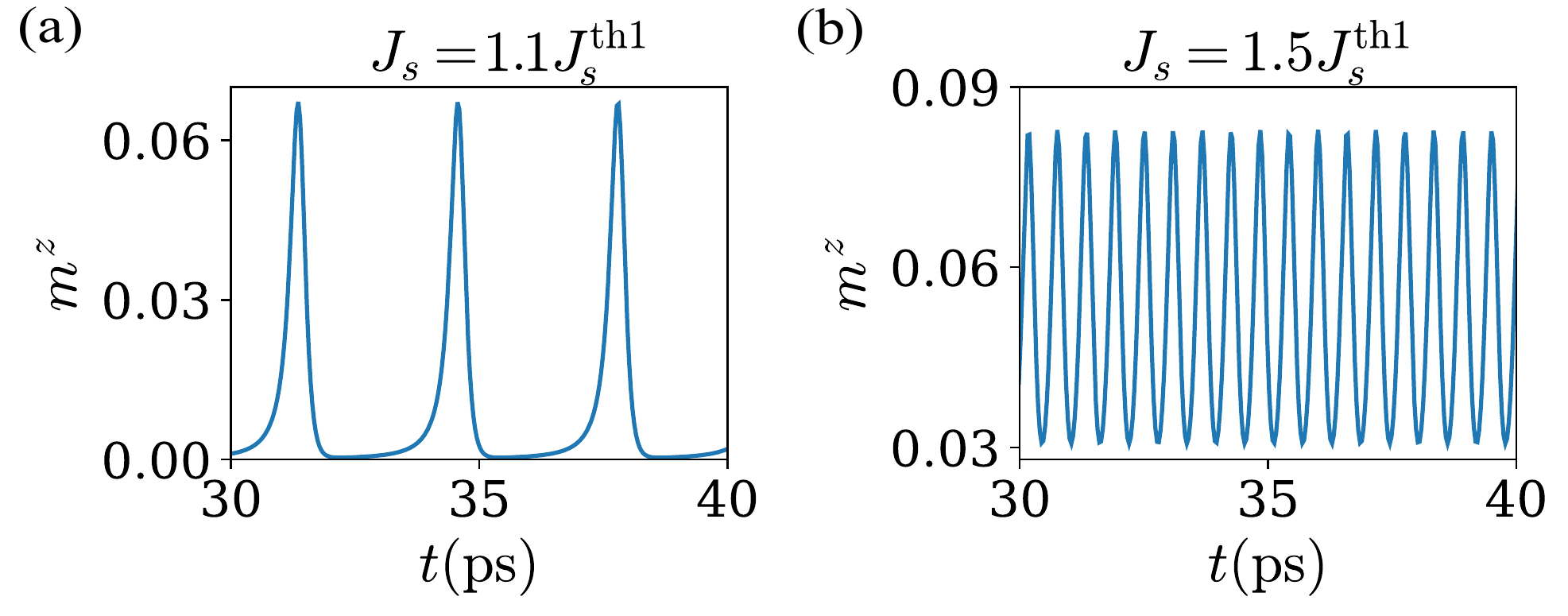}
  \caption{${m}^{z}$ for larger damping, $\alpha = 0.1$, and $d_{a} = 4 \ \mathrm{nm}$. (a) Non-coherent \textcolor{black}{(spike-like)} signals \textcolor{black}{near} the threshold current $J_{s} = 1.1J_{s}^{\mathrm{th1}}$. (b) Coherent signal for larger current $J_{s} = 1.5J_{s}^{\mathrm{th1}}$. The angular frequency is directly proportional to $m^{z}$, and therefore it would show the exact same features (in the absence of any external field) for the chosen values of current.} 
  \label{fig:spikes}
\end{figure}

Metallic AFMs such as $\mathrm{Mn_{3}Ir}$ and $\mathrm{Mn_{3}Sn}$ could be considered as examples of $+\pi/2$ and $-\pi/2$ chiralities, respectively. Recently, thin-films with different thickness ranging from $1 \ \mathrm{nm}$ to $5 \ \mathrm{nm}$ of both these materials have been grown using UHV magnetron sputtering~\citep{reichlova2015current, markou2018noncollinear, taylor2019magnetic, siddiqui2020metallic}.
In addition, different values of damping constants have been reported for $\mathrm{Mn_{3}Sn}$~\citep{yamane2019dynamics, nyari2019weak}.
Therefore, we expect the results presented in  Sections~\ref{comparison},~\ref{extract} and~\ref{micro} to be useful for benchmarking THz dynamics in experimental set-ups with such thin films metallic antiferromagnets. 



\begin{figure}[ht!]
  \centering
  \includegraphics[width = \columnwidth, clip = true, trim = 0mm 2mm 0mm 0mm]{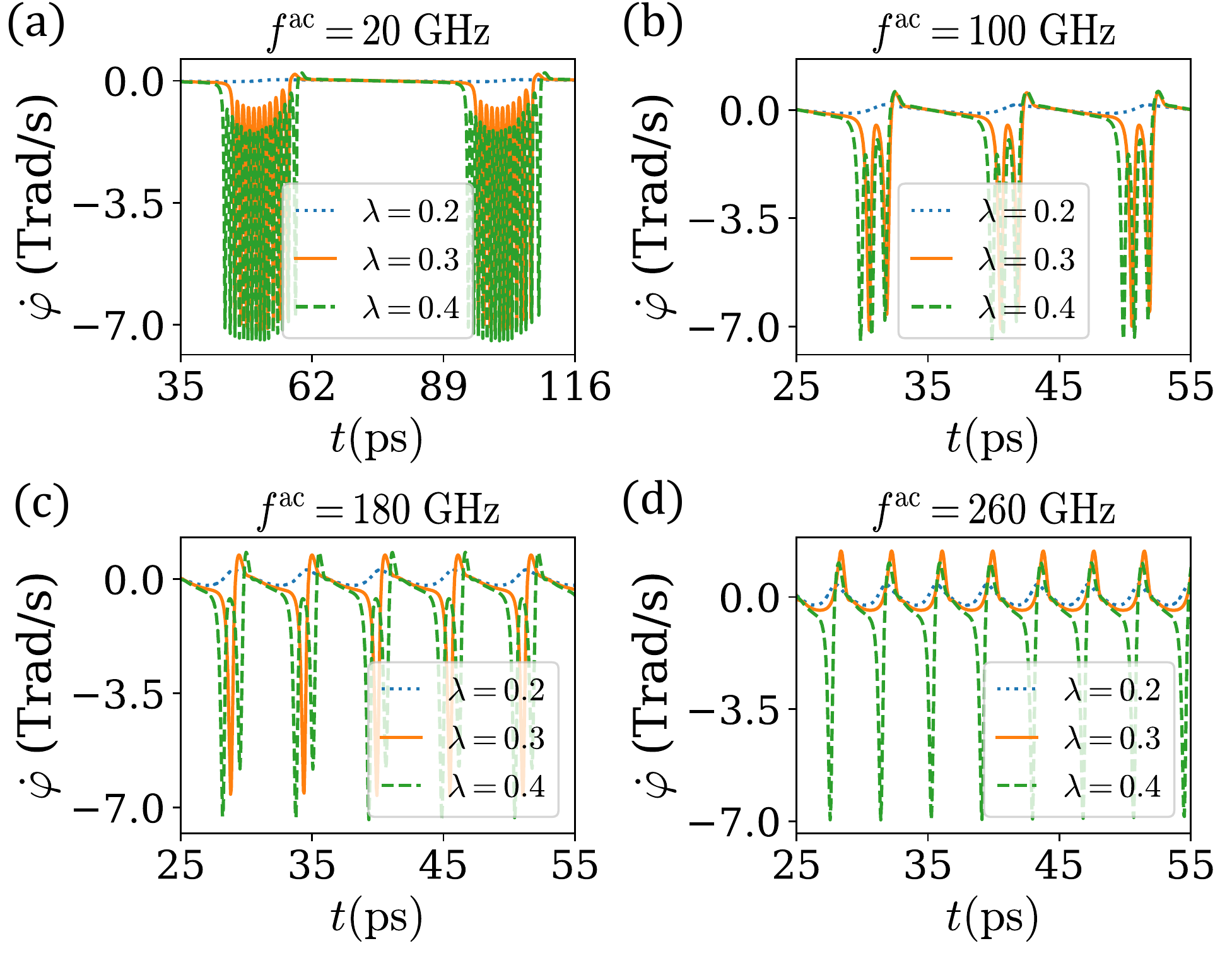}
  \caption{Time dynamics (single and train of spikes) of a single ``neuron'' for different input currents and frequencies. The net input current should be greater than the threshold current ($\lambda > 0.2$) for a non-zero dynamics. For an input current above the threshold, as the external frequency increases the dynamics changes from (a) bursts of spikes to (c) single spikes to (d) no spikes ($\lambda = 0.3$). As the input current increases to $\lambda = 0.4$ the range of external frequency where the spiking behaviour is observed increases.} 
  \label{fig:neuron1}
\end{figure}
\vspace{-10pt}
\section{Potential Applications}\label{appl}
Neurons in the human brain could be thought of as a network of coupled non-linear oscillators, while the stimuli to excite neuronal dynamics is derived from the neighboring neurons in the network~\citep{grollier2016spintronic, roy2019towards, grollier2020neuromorphic, kurenkov2020neuromorphic}.
For materials with $+\pi/2$ chirality, a non-linear behaviour was observed for large damping, and input currents near the threshold current, $J_{s}^{\mathrm{th1}}$, in Fig.~\ref{fig:freq}(b), (f). 
This non-linearity corresponds to Dirac-comb-like magnetization dynamics, as shown in Fig.~\ref{fig:spikes}(a), and is similar to the dynamics of biological neurons in their spiking behaviour as well as a dependence on the input threshold. 
However, unlike a biological neuron which shows various dynamical modes such as spiking, bursting, and chattering~\citep{izhikevich2003simple}, the dynamics here shows only spikes and does not show any refractory (``resting'') period. 
Recent works~\citep{sulymenko2018ultra, khymyn2018ultra} have shown that it is possible to generate single spiking as well as bursting behaviours using NiO-based AFM oscillators by considering an input DC current below $J_{s}^{\mathrm{th1}}$, and superimposing it with an AC current. As the AC current changes with time, the total current could either go above the threshold, thereby triggering a non-linear response, or below the threshold current resulting in a ``resting'' period. 
Here we explore the possibility of spiking behaviours in $+\pi/2$ chirality materials such as Mn$_3$Ir under the effect of an input spin current.
We use the non-linear pendulum model of Eq.~(\ref{eq:pend1}) and study the possible dynamics in case of a single oscillator, two unidirectional coupled oscillators, and two bidirectional coupled oscillators.
\vspace{-5pt}
\subsection{Ultra-fast Hardware Emulator of Neurons}
We consider a large damping of $\alpha = 0.1$ while the other material parameters correspond to that of Mn$_{3}$Ir as listed in Table~\ref{tab:materials}. Next we choose an input current $J_{s} (t) = J_{s}^{\mathrm{dc}} + J_{s}^{\mathrm{ac}} (t)$, where $J_{s}^{\mathrm{dc}} = 0.8 J_{s}^{\mathrm{th1}}$ is the dc component of the input current, superimposed with a smaller ac signal $J_{s}^{\mathrm{ac}} (t) = \lambda J_{s}^{\mathrm{th1}} \cos(2 \pi f^{\mathrm{ac}} t)$. The time dynamics of this non-linear oscillator is governed by 
\begin{equation}\label{eq:neuron1}
    \ddot{\varphi} + \alpha \omega_{E} \dot{\varphi} + \omega_{E} \frac{\omega_{K}}{2} \sin{2\varphi} + \omega_{E} \omega_{s} (t) = 0,
\end{equation}
where $\omega_{s} (t) (\propto J_s(t))$ is the time varying input current. 

\begin{figure}[ht!]
  \centering
  \includegraphics[width = \columnwidth, clip = true, trim = 0mm 2mm 0mm 0mm]{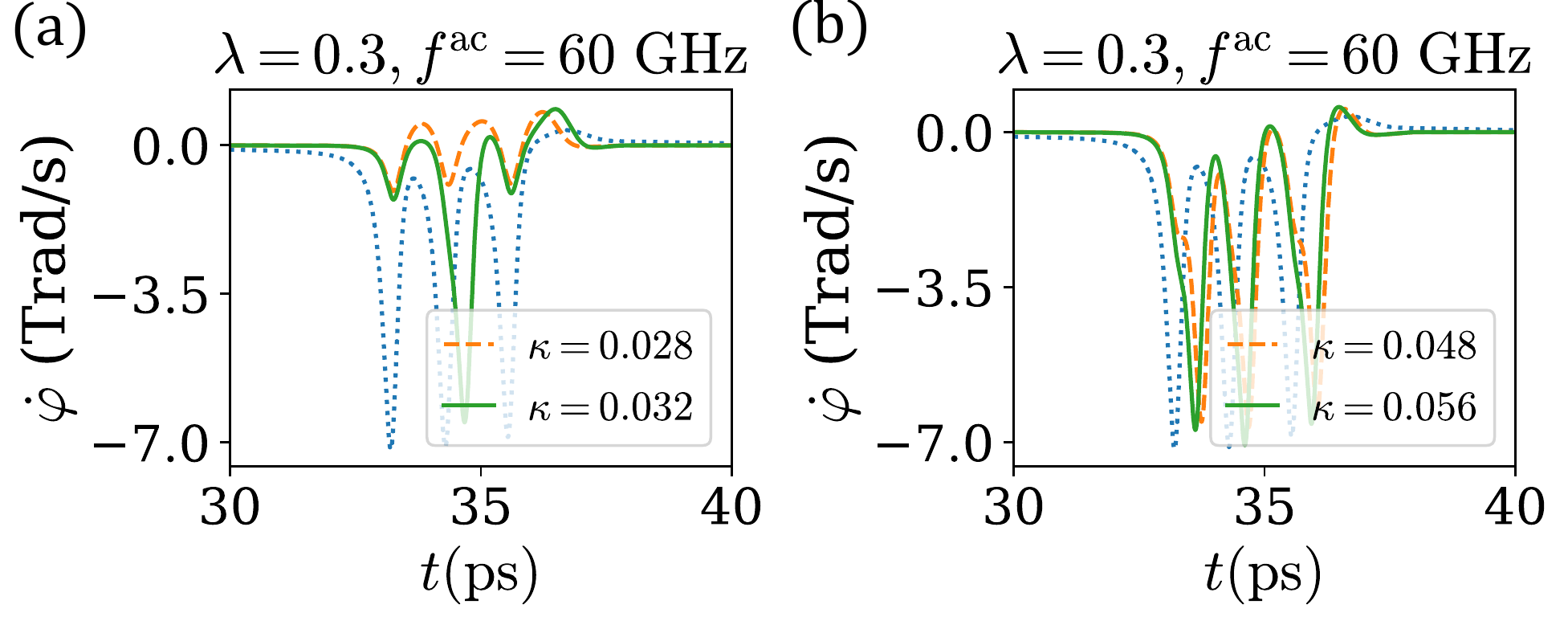}
  \caption{The dynamics of two neuron system with unidirectional coupling at $f^{\mathrm{ac}} = \textcolor{black}{60}$ GHz, and $\lambda = 0.3$. The dotted blue curve corresponds to the first neuron. (a) Second neuron shows no spike for $\kappa = 0.028$ but a single spike for $\kappa = 0.032$. (b) The single spiking behaviour changes to bursts with three spikes as $\kappa$ increases and coupling strengthens.} 
  \label{fig:uni_coupled60}
\end{figure}

Figure~\ref{fig:neuron1} presents the dynamics of Eq.~(\ref{eq:neuron1}) for different input current and frequencies.  
Firstly, it can be observed that the input current must be greater than the threshold current to excite any dynamics viz. $\lambda$ must be greater than 0.2 (dotted line corresponding to $\lambda = 0.2$ shows no spikes for any value of external frequency). 
Secondly, for input currents above the threshold viz. $\lambda = \{0.3, 0.4\}$, a train of spikes is observed for lower frequency of 20 GHz in Fig.~\ref{fig:neuron1}(a). 
However, as frequency of the input excitation increases the number of observed spikes decreases for both values of current considered here (Fig.~\ref{fig:neuron1}(b, c)).
Finally, it can be observed from Fig.~\ref{fig:neuron1}(d) that for very large frequency the spiking behaviour vanishes for lower current ($\lambda = 0.3$) but persists for higher current ($\lambda = 0.4$). For higher values of current, the cut-off frequency is higher.
This observed spiking behaviour is indeed similar to that of biological neurons~\citep{izhikevich2003simple}. Here, however, the observed dynamics is very fast in the THz regime and thus the AFM oscillators could be used as the building blocks of an ultra-high throughput brain-inspired computing architecture. 

\begin{figure}[ht!]
  \centering
  \includegraphics[width = \columnwidth, clip = true, trim = 0mm 0mm 0mm 0mm]{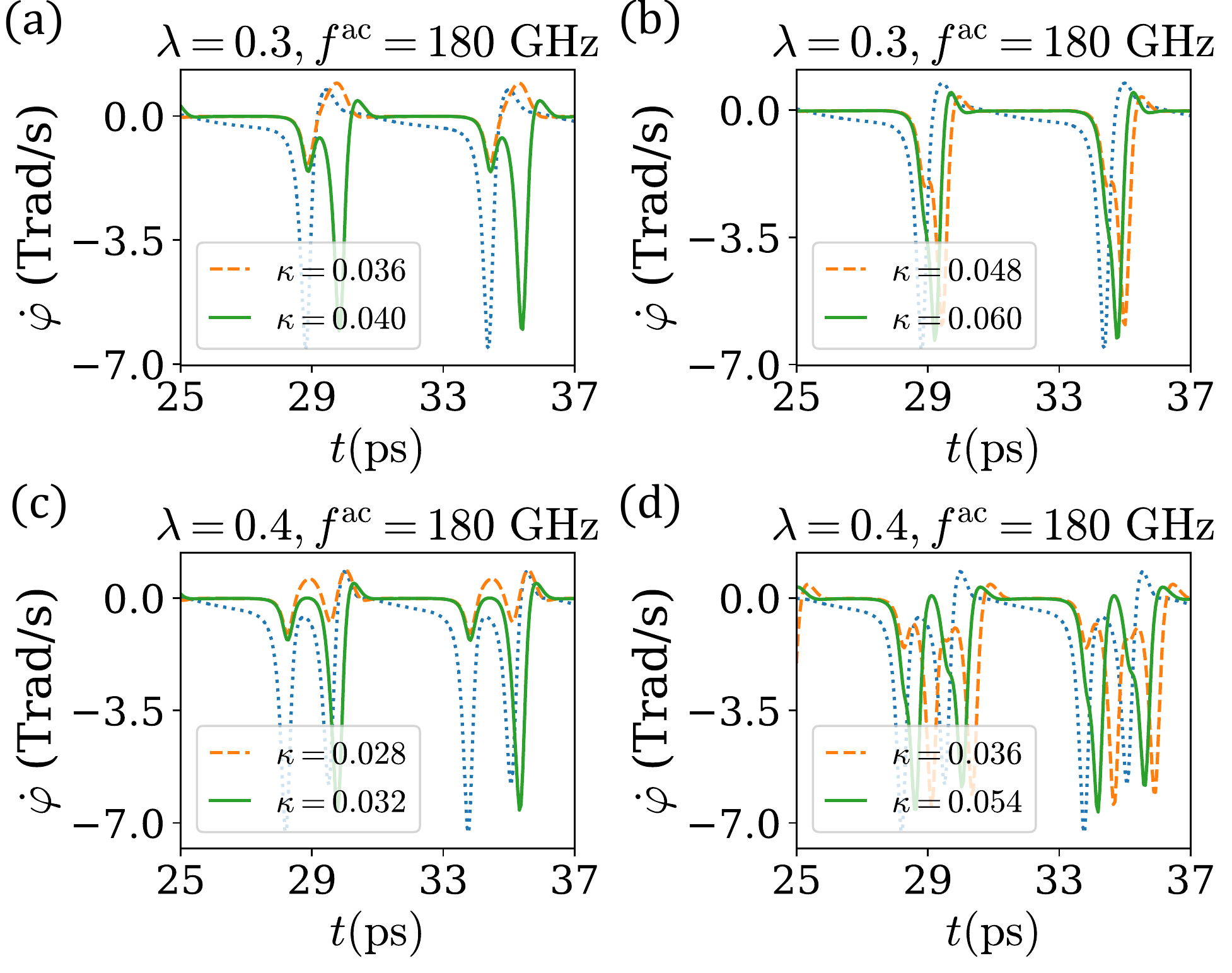}
  \caption{The dynamics of two neuron system with unidirectional coupling at $f^{\mathrm{ac}} = \textcolor{black}{180}$ GHz. The dashed blue curve corresponds to the first neuron. $\lambda = 0.3$: (a) Single spike for $\kappa = 0.04$ but not for $\kappa = 0.036$. (b) The single spiking behaviour become prominent as the coupling strengthens. $\lambda = 0.4$: (c) Single spike for $\kappa = 0.032$ in response to a double spiking behaviour of the first neuron. (d) For larger $\kappa$ second neuron shows bursting dynamics with two spikes.} 
  \label{fig:uni_coupled180}
\end{figure}
\subsection{Two unidirectional coupled artificial neurons}
A network composed of interacting oscillators forms the basis of the oscillatory neurocomputing model proposed by Hoppensteadt and Izhikevich~\citep{hoppensteadt1999oscillatory}.
In such a network, the dynamics of an oscillating neuron (or a ``node") is controlled by the incoming input signal as well as its coupling to neighboring neurons. 
To investigate this coupling behaviour we consider a system of two unidirectional coupled neurons. The first neuron is driven by an external signal and its dynamics is governed by Eq.~(\ref{eq:neuron1}). The dynamics of the second neuron, on the other hand, depends on the output signal of the first neuron as well as the coupling between the two neurons. It is governed by 
\begin{align}\label{eq:neuron2}
    \begin{split}
        \ddot{\varphi}_{j} + \alpha \omega_{E} \dot{\varphi}_{j} + \omega_{E} \frac{\omega_{K}}{2} \sin{2\varphi_{j}} &+ \omega_{E} \omega_{s} \\
        &- \kappa_{ij} \omega_{E} \dot{\varphi}_{i} sgn(\omega_{s})= 0,
    \end{split}
\end{align}
where $\kappa_{ij} = \kappa$ is the unidirectional coupling coefficient from neuron $i = 1$ to $j = 2$. 
There is no feedback from the second neuron to the first and therefore $\kappa_{ji} = 0$. 
In addition to the input from the first neuron, the second neuron is also driven by a constant DC current $J_{s2}^{\mathrm{dc}}$ ($\propto \omega_s$ in Eq.~(\ref{eq:neuron2})). We choose this DC current to be the same as that for the first neuron viz. $J_{s2}^{\mathrm{dc}} = 0.8 J_{s}^{\mathrm{th1}}$. 
The dynamics of the second neuron for two different external input currents $(\lambda = \{0.3, 0.4\})$ and frequencies ($f^{\mathrm{ac}} = \{60, 180\}$ GHz) is presented in Figs.~\ref{fig:uni_coupled60} and~\ref{fig:uni_coupled180}. 

Firstly, it can be observed that in all the cases the second neuron shows a spiking behaviour only for $\kappa$ above a certain value. 
Secondly, for $\lambda = 0.3$ and $f^{\mathrm{ac}} = 60$ GHz, wherein the first neuron shows bursting behaviour consisting of three spikes, the second neuron shows a single spike (Fig.~\ref{fig:uni_coupled60}(a)) for lower value of $\kappa$, and three spikes for stronger coupling (Fig.~\ref{fig:uni_coupled60}(b)). 
This behaviour is due to the threshold dependence of the second neuron as well as due to its inertial dynamics.
Similar behaviour is also observed for $\lambda = 0.4$, and $f^{\mathrm{ac}} = 180$ GHz in Figs.~\ref{fig:uni_coupled180}(c), (d).
Thirdly, for $\lambda = 0.3$ and $f^{\mathrm{ac}} = 180$ GHz, wherein the first neuron shows a single spike, Fig.~\ref{fig:uni_coupled180}(a) shows that compared to the case of $f^{\mathrm{ac}} = 60$ GHz a slightly higher value of $\kappa$ is now required to excite the second neuron. 
The single spiking behaviour of the second neuron becomes more prominent as the coupling strength increases because of a stronger input as shown in Fig.~\ref{fig:uni_coupled180}(b). 
Recently, it was suggested that this coupled behaviour of THz artificial neurons could be used to build ultra-fast multi-input AND, OR, and majority logic gates~\citep{sulymenko2018ultra}. 

\begin{figure}[ht!]
  \centering
  \includegraphics[width = \columnwidth, clip = true, trim = 0mm 2mm 0mm 0mm]{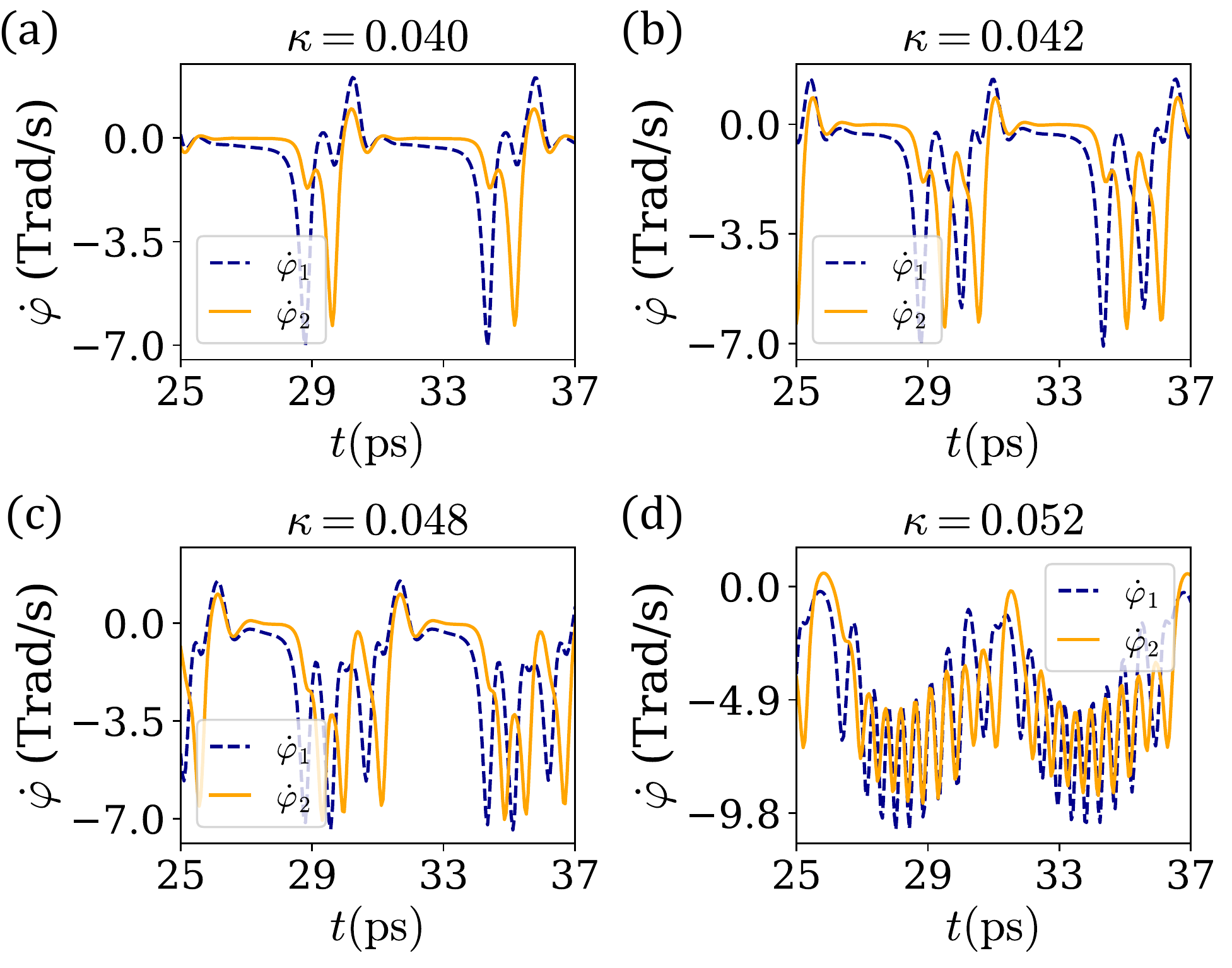}
  \caption{The dynamics of two neuron system with bidirectional coupling at $f^{\mathrm{ac}} = \textcolor{black}{180}$ GHz, and $\lambda = 0.3$. First neuron shows bursting behaviour in this system while the second neuron follows the first neuron for all values of $\kappa$. As the coupling between the two neurons increase the number of spikes for both the neurons increases.} 
  \label{fig:bi_coupled180_0.3}
\end{figure}
\subsection{Two bidirectional coupled artificial neurons}
\begin{figure}[ht!]
  \centering
  \includegraphics[width = \columnwidth, clip = true, trim = 0mm 2mm 0mm 0mm]{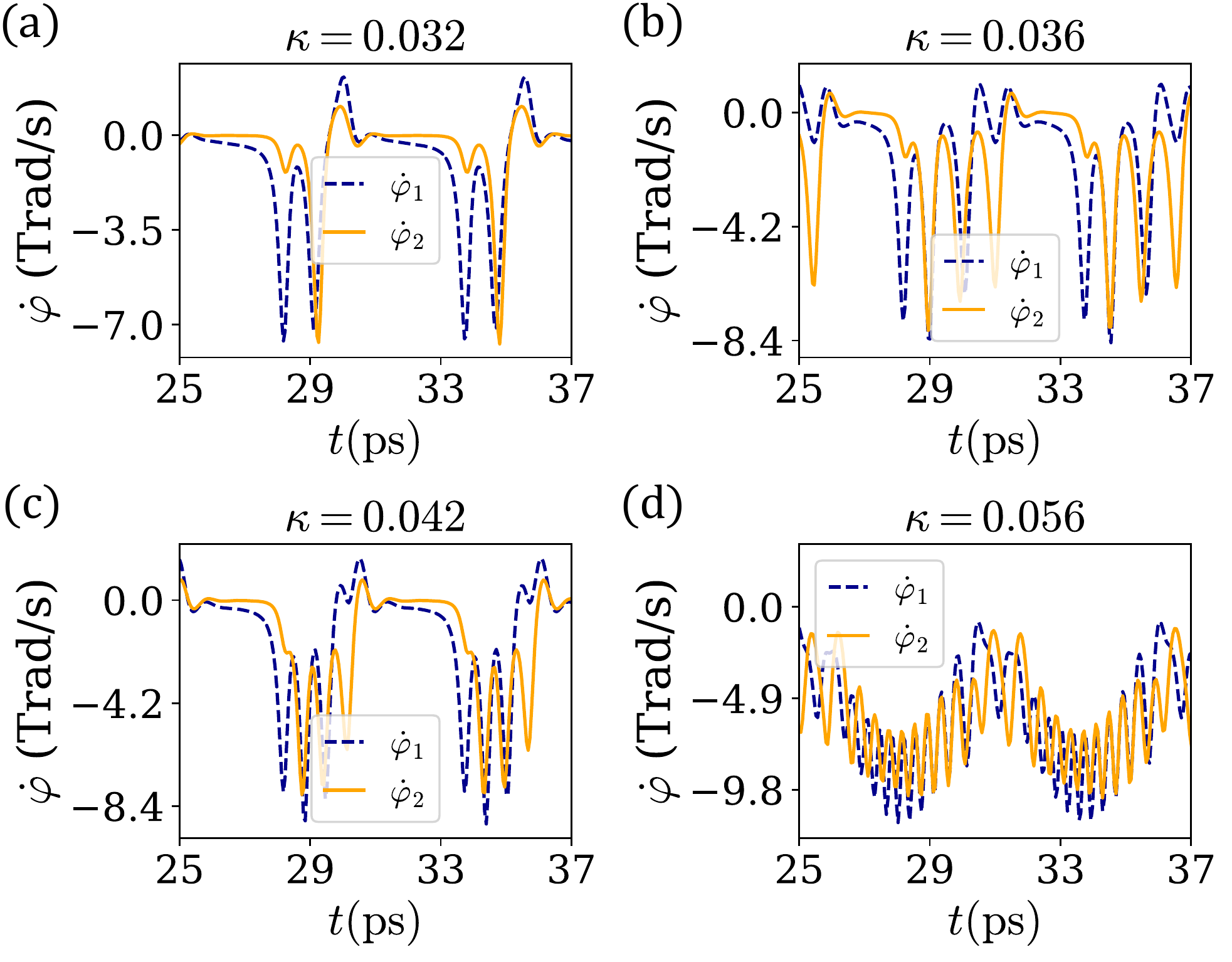}
  \caption{The dynamics of two neuron system with bidirectional coupling at $f^{\mathrm{ac}} = \textcolor{black}{180}$ GHz, and $\lambda = 0.4$. Second neuron fires when the coupling is above a certain threshold which in turn leads to another spike for the first neuron. As the coupling between the two neurons increase the number of spikes for both the neurons increases.} 
  \label{fig:bi_coupled180_0.4}
\end{figure}

In some circuits it is possible that the coupling between any two neurons is bidirectional. In such cases, in addition to a forward coupling from the first neuron to the second, a feedback exists from the second neuron to the first. The dynamics of each neuron of this coupled system is governed by Eq.~(\ref{eq:neuron2}), however, $\omega_{s} = \omega_{s} (t)$ for the first neuron, as discussed previously. We consider $\kappa_{12} = \kappa_{21} = \kappa$. Figures~\ref{fig:bi_coupled180_0.3} and~\ref{fig:bi_coupled180_0.4} show the dynamics of the two neurons of this coupled system with the coupling $\kappa$ at $f^{\mathrm{ac}} = \textcolor{black}{180}$ GHz for $\lambda = 0.3$ and 0.4, respectively. 

Firstly, \textcolor{black}{Fig.~\ref{fig:bi_coupled180_0.3}(a) shows that for $\kappa = 0.04$ the dynamics of both $\dot{\varphi}_{1}$ and $\dot{\varphi}_{2}$ are almost similar to that presented in Fig.~\ref{fig:uni_coupled180}(a), viz. the effects of coupling is very small. However, as the coupling between the two neurons increases (Fig.~\ref{fig:bi_coupled180_0.3}(b)-(d)), a positive feedback is established between the two neuron leading to dynamics with two or more spikes, in general. This is observed after the second neuron has fired, at least once, because the positive feedback leads to a net input greater than the threshold current to the first neuron, even though the external input has reduced below the threshold. Similar behavior is also observed in the case of $\lambda = 0.4$, although at lower values of coupling, as presented in Fig.~\ref{fig:bi_coupled180_0.4}. 
The results allude to the threshold behaviour of the neurons, inertial nature of the dynamics, and a dependence of the dynamics on the phase difference between the two neurons.}
The dynamics of two bidirectional coupled artificial neurons presented here could be the first step towards building AFM-based recurrent neural networks or reservoir computing~\citep{csaba2020coupled}, instead of the slower FM-based coupled oscillator systems~\citep{torrejon2017neuromorphic, romera2018vowel}.


\section{Conclusion}
In this work, we numerically and theoretically explore the THz dynamics of thin-film metallic non-collinear coplanar AFMs such as $\mathrm{Mn_{3}Ir}$ and $\mathrm{Mn_{3}Sn}$, under the action of an injected spin current with spin polarization perpendicular to the plane of the film. 
Physically, these two AFM materials differ in their spin configuration viz. positive chirality for Mn$_3$Ir, and negative chirality for Mn$_3$Sn.
In order to explore the dynamics numerically, we solve three LLG equations coupled to each other via inter-sublattice exchange interactions. 
We also analyze the dynamics theoretically in the limit of strong exchange and show that it can be mapped to that of a damped-driven pendulum if the effects of inhomogeneity in the material are ignored. 
We find that the dynamics of Mn$_3$Ir is best described by a non-linear pendulum equation and has a hysteretic behaviour, while that of Mn$_3$Sn \textcolor{black}{in the THz regime} is best described by a linear pendulum equation and has \textcolor{black}{a significantly small} threshold for oscillation.
The hysteretic dynamics in the case of Mn$_3$Ir allows for possibility of energy efficient THz coherent sources. 
On the other hand, \textcolor{black}{a small} threshold current requirement in the case of Mn$_3$Sn indicates the possibility of efficient coherent signal sources from MHz to THz regime. 
We employ the TAMR detection scheme to extract the THz oscillations as time-varying voltage signals across an external resistive load.
Including inhomogeneous effects leads to a variation in the dynamics --- the lower threshold current for sustaining the dynamics increases, the hysteretic region reduces, and the frequency of oscillation decreases for lower current levels.
Finally, we also show that the non-linear behaviour of positive chirality materials with large damping could be used to emulate artificial neurons. 
An interacting network of such oscillators could enable the development of neurocomputing circuits for various cognitive tasks.
The device setup and the results presented in this paper should be useful in designing experiments to further study and explore THz oscillations in thin-film metallic AFMs.

\vspace{5pt}
\section*{acknowledgements}
This research is funded by AFRL/AFOSR, under AFRL Contract No. FA8750-21-1-0002. The authors also acknowledge the support of National Science Foundation through the grant no. CCF-2021230.
Ankit Shukla is also grateful to Siyuan Qian for fruitful discussions.

\nocite{*}
\bibliography{apssamp}

\end{document}